\documentclass[10pt,twocolumn]{article}

\usepackage[letterpaper,margin=0.75in]{geometry}
\usepackage{mathptmx}
\usepackage[T1]{fontenc}
\usepackage[utf8]{inputenc}
\usepackage[kerning,spacing]{microtype}

\usepackage{nicefrac}
\usepackage{siunitx}
\usepackage{array,framed}
\usepackage{booktabs}
\usepackage{
  color,
  float,
  wrapfig,
  graphicx,
  subcaption
}
\usepackage{textcomp}
\usepackage{setspace}
\usepackage{latexsym,fancyhdr,url}
\usepackage{xurl}
\usepackage{enumerate}
\usepackage{xparse}
\usepackage{xspace}
\usepackage{multirow}
\usepackage{csvsimple}
\usepackage{balance}
\usepackage{tikz}
\usepackage{amssymb}
\usepackage{adjustbox}
\usepackage{hyperref}
\usepackage{tabularx}
\usepackage{listings}
\usepackage{pdfpages}
\usepackage{amsmath}

\usepackage{amsthm}
\usepackage[labelfont=sc,textfont=normalfont]{caption}
\usepackage{xcolor}
\usepackage{colortbl}
\usepackage{longtable}
\usepackage{changepage}
\usepackage{fvextra}
\usepackage{enumitem}
\usepackage{mathrsfs}
\usepackage[most]{tcolorbox}
\usepackage{algorithm}
\usepackage{algpseudocode}
\usepackage{placeins}
\usepackage{dblfloatfix}
\usepackage{mathtools}

\definecolor{timeoutRed}{HTML}{E88A8A}
\definecolor{tteGreen01}{HTML}{489955}
\definecolor{tteGreen02}{HTML}{4FA85E}
\definecolor{tteGreen03}{HTML}{5BB26A}
\definecolor{tteGreen04}{HTML}{6AB977}
\definecolor{tteGreen05}{HTML}{79C085}
\definecolor{tteGreen06}{HTML}{88C793}
\definecolor{tteGreen07}{HTML}{98CEA1}
\definecolor{tteGreen08}{HTML}{A7D5AF}
\definecolor{tteGreen09}{HTML}{B6DDBC}
\definecolor{tteGreen10}{HTML}{C5E4CA}
\definecolor{tteGreen11}{HTML}{D4EBD8}
\definecolor{tteGreen12}{HTML}{E3F2E6}

\lstdefinestyle{diglisting}{
  basicstyle=\ttfamily\footnotesize\mdseries,
  numbers=left,
  numberstyle=\tiny\color{black!55}\mdseries,
  stepnumber=1,
  numbersep=6pt,
  frame=none,
  breaklines=true,
  showstringspaces=false,
  tabsize=2,
  commentstyle=\color{orange!75!red}\mdseries\itshape,
  keywordstyle=\color{blue!70!black}\bfseries,
  stringstyle=\color{green!45!black}\mdseries,
  aboveskip=3pt,
  belowskip=3pt
}
\newtheoremstyle{definitionstyle}
  {6pt}
  {6pt}
  {\itshape}
  {}
  {\scshape}
  {.}
  {.5em}
  {}
\theoremstyle{definitionstyle}
\newtheorem{definition}{Definition}

\hypersetup{
    colorlinks=true,
    linkcolor=blue,
    filecolor=blue,
    urlcolor=blue,
}

\newcommand*\circled[1]{\tikz[baseline=(char.base)]{
            \node[shape=circle,fill,inner sep=0.3pt] (char) {\small\textcolor{white}{#1}};}}

\usepackage{
  tikz,
  pgfplots,
  pgfplotstable
}
\usetikzlibrary{
  shapes.geometric,
  arrows,
  external,
  pgfplots.groupplots,
  matrix
}
\pgfplotsset{compat=1.9}

\DeclareMathAlphabet{\mathcal}{OMS}{cmsy}{m}{n}

\DeclareGraphicsExtensions{%
    .png,.PNG,%
    .pdf,.PDF,%
    .jpg,.mps,.jpeg,.jbig2,.jb2,.JPG,.JPEG,.JBIG2,.JB2}

\def\Snospace~{\S{}}

\makeatletter
\renewcommand\subsubsection{\@startsection{subsubsection}{3}{\z@}%
                          {-3.25ex\@plus -1ex \@minus -.2ex}%
                          {-0.8em}%
                          {\normalfont\normalsize\itshape}}
\makeatother
\begin{document}

\title{DIG: Oracle-Guided Directed Input Generation for One-Day Vulnerabilities}
\author{
Andrew Bao\\
University of Minnesota, Twin Cities\\
\texttt{bao00065@umn.edu}
\and
Haochen Zeng\\
University of California, Riverside\\
\texttt{hzeng013@ucr.edu}
\and
Peng Chen\\
Independent Researcher\\
\texttt{spinpx@gmail.com}
\and
Stephen McCamant\\
University of Minnesota, Twin Cities\\
\texttt{smccaman@umn.edu}
\and
Pen-Chung Yew\\
University of Minnesota, Twin Cities\\
\texttt{yew@umn.edu}
}
\date{}

\maketitle
\raggedbottom

\begin{abstract}
One-day vulnerabilities pose significant risks due to delayed or incomplete patch adoption. 
Generating proof-of-concept (PoC) inputs is therefore essential for assessing real-world impact.
The key challenge is identifying necessary constraints for triggering the vulnerability 
and solving them effectively.
Existing directed fuzzing approaches prioritize inputs toward target locations, 
but neither explicitly identify necessary constraints nor solve them effectively, 
relying instead on target-distance feedback and random mutation.
Agentic approaches show strong potential through code reasoning and structured input generation,
but goal drift in long-horizon reasoning limits their effectiveness.

DIG addresses this challenge by exploiting a key property of one-day vulnerabilities: patches
often reveal necessary preconditions for triggering. 
DIG uses an LLM to analyze the patch and synthesize an oracle making these conditions explicit.
The oracle supports effective PoC generation at two levels. 
At the high level, DIG performs oracle-guided generator evolution, 
where an agent infers and solves constraints to satisfy the oracle. 
At the low level, DIG instruments the oracle into the target program and uses branch-distance feedback to 
guide random mutation in directed fuzzing.
Evaluation shows DIG outperforms 2 state-of-the-art 
agents and 10 fuzzers across 138 real-world CVEs. DIG triggers 80 vulnerabilities, 
surpassing prior results and outperforming the best baseline by 40\% 
(57 vs. 80 CVEs). Notably, DIG exclusively triggers 9 vulnerabilities no 
existing technique can trigger. Compared to the average of other tools, 
DIG triggers vulnerabilities faster in 92.9\% of cases, achieving over 100$\times$ speedup in 48.8\% of cases, 
with a maximum speedup of 3,664$\times$.
Beyond one-day PoC generation, DIG uncovers 6 previously 
unknown vulnerabilities in widely deployed libraries, enabling zero-day discovery.
\end{abstract}

\section{Introduction}
\label{sec:intro}
Modern software systems have become foundational infrastructure for the internet and
critical services.
As these systems scale and become more complex, security vulnerabilities have become
increasingly prevalent and difficult to assess.
The vulnerability lifecycle distinguishes two critical periods: zero-day vulnerabilities 
exist from discovery until patch release, while \emph{one-day vulnerabilities} exist from 
patch disclosure until widespread deployment.
Despite the fact that zero-day vulnerabilities receive significant attention, one-day 
vulnerabilities present an equally severe but often underestimated threat.

Prior studies~\cite{tan2021locating,dissanayake2022and,dissanayake2022empirical,zimmermann2019small,vasilakis2018breakapp,xiao2025jbomaudit,williams2025research}
show that patch adoption is often slow and incomplete.
During this window, attackers can analyze released patches to infer vulnerable behavior
and rapidly develop exploits. 
From a defender's perspective,
the availability of a patch alone is insufficient to assess the risk of one-day vulnerabilities. 
Defenders must determine whether their specific systems remain vulnerable under 
heterogeneous deployment conditions~\cite{nappa2015attack}, assess the severity of 
potential exploitation~\cite{xiao2018patching}, and verify patch 
completeness~\cite{li2017large,wu2021feasibility,wang2025cybergym}.
This attack--defense asymmetry renders one-day vulnerabilities, in practice, nearly as dangerous as zero-days.
To bridge this gap, defenders need automated proof-of-concept (PoC) generation that 
provides end-to-end evidence of real-world risk.

The key challenge of PoC generation for one-day vulnerabilities is twofold:
identifying which constraints are necessary for vulnerability triggering,
and solving these constraints effectively.
Existing approaches tackle this challenge from two different directions.

First, directed greybox fuzzing (DGF) does not 
explicitly identify which constraints are necessary for vulnerability triggering. Instead, it identifies target locations 
(e.g., vulnerable function locations) and uses control-flow~\cite{bohme2017directed,du2022windranger} or
data-flow~\cite{luo2023selectfuzz,kim2023dafl} distances to estimate whether an input
is making progress toward these targets.
Because fuzzing provides high-throughput execution, DGF can explore a large number of
candidate inputs and gradually select inputs that make progress toward the targets. 
However, since this guidance is only an indirect proxy for constraints that are necessary for vulnerability triggering,
and the search still relies primarily on low-level \emph{random mutation},
DGF often struggles to construct valid PoCs.

Second, recent LLM-based input generation approaches~\cite{zhang2025low,meng2024large,chen2025elfuzz} offer a promising direction. 
Since LLMs are trained on large-scale code, they capture broad knowledge about program syntax, input formats, and common semantic rules.
PBFuzz~\cite{zeng2025pbfuzz} further strengthens these approaches through an agentic workflow.
To identify constraints, the agent analyzes the target code, call relationships, and execution feedback
to infer reachability and triggering constraints.
To solve these constraints, it translates them into parameterized input generators,
and then uses property-based testing to systematically search for inputs that satisfy them.
However, many hard-to-trigger vulnerabilities remain difficult for PBFuzz. 
This is because PoC generation is often a long-horizon reasoning task that requires satisfying 
a sequence of constraints to trigger the vulnerability.
As a result, agents are prone to the \emph{goal drift} effect~\cite{wang2026reasoning,wang2026long,arike2025technical}, 
where they may gradually lose sight of the vulnerability trigger over long
reasoning chains, causing them to infer incorrect constraints, forget valid ones,
or fail to preserve the constraints necessary for constructing a valid PoC.

These limitations suggest that effective one-day PoC generation requires both a more direct source of information
for identifying the constraints necessary for vulnerability triggering and a more effective way to solve them.
Our key insight comes from a distinctive property of one-day vulnerabilities: A security patch may add a 
new check, correct a variable relation, introduce a state constraint, or modify a precondition before a dangerous operation. 
While these changes appear to fix the vulnerability, they more fundamentally reveal 
why the original vulnerability could be triggered. 
Therefore, a patch is not only the result of a fix, but also an important source for understanding the 
constraints required to trigger the vulnerability.
For example, a patch that adds a bounds check, \texttt{if (size <= MAX)}, directly 
suggests that triggering the vulnerability may require violating this check, i.e., \texttt{size > MAX}.

We derive an \emph{oracle} from the patch and define it as a necessary precondition for vulnerability triggering.
The \emph{oracle} supports effective PoC generation at two levels.
At the high level, the oracle serves as an explicit starting point for identifying constraints.
Starting from the oracle, the agent reasons backward to infer a sequence of constraints that lead to the oracle.
The derived constraints give each iteration in the agentic workflow a concrete subgoal: solving unresolved constraints and
gradually building the PoC, which helps mitigate the \emph{goal drift} problem.
At the low level, the oracle is instrumented into the binary.
This gives the directed fuzzer explicit constraints to solve, rather than relying only on indirect target-distance guidance.
Moreover, the oracle makes constraint solving more effective by improving \emph{random mutation} with branch-distance feedback.

\paragraph{Our solution}
We present DIG (\textbf{D}irected \textbf{I}nput \textbf{G}eneration), an oracle-guided PoC generation system for one-day vulnerabilities.
Specifically, DIG introduces the following techniques:

\begin{itemize}[leftmargin=*, itemsep=3pt, topsep=3pt]
    \item \textbf{Oracle Synthesis:} LLMs are well suited for patch analysis because, through training on 
    large-scale code and bug-fix patterns, they can understand the 
    semantic intent of a patch rather than merely its textual changes. 
    DIG leverages this capability to turn implicit patch information into an oracle. 
    First, DIG identifies the key constraints in the patch that are most relevant to the vulnerability, 
    such as a newly added check, a corrected variable relation, or an enforced state condition. 
    Second, it analyzes how these constraints prevent the original vulnerable behavior in the
    context of the source code and vulnerability description. 
    Third, it infers what must hold in the vulnerable source code for 
    these constraints to be violated, which DIG uses as the oracle.

\item \textbf{Oracle-guided Generator Evolution:}
    At the high level, DIG's agent performs LLM-based code reasoning from the oracle to infer the constraints that 
    govern how variables in the oracle are computed and propagated. 
    DIG then evolves the generator based on the inferred constraints: 
    each run preserves the generation logic developed in the previous run and
    extends it to address additional unresolved constraints. 
    This incremental refinement breaks the goal of PoC generation into a sequence of smaller constraint-solving steps, 
    mitigating \emph{goal drift} in agent reasoning and progressively improving the generator's ability to
    produce valid PoCs.

\item \textbf{Oracle-guided Directed Mutation:}
    At the low level, DIG instruments the oracle into the binary, which allows DIG to improve \emph{random mutation} with 
    oracle guidance. 
    Mutations are concentrated on input regions that directly influence oracle satisfaction,
    while branch-distance feedback steers the search direction, guiding the fuzzer to move 
    the input closer to satisfying the oracle rather than further away.
    We argue that this low-level directed mutation complements high-level generator evolution.
    While high-level reasoning is effective at global search and constructing inputs that satisfy
    the constraints leading to the oracle, it is relatively slow and may still
    leave the oracle unsatisfied (e.g., hard-to-infer constraints due to hidden program states). 
    Low-level directed mutation complements this process by performing high-throughput
    local search over small changes to oracle-relevant input regions.

\end{itemize}

\paragraph{Contributions}
In summary, we make the following contributions:
\begin{itemize}[leftmargin=*, itemsep=3pt, topsep=3pt]
\item At the conceptual level, we formulate one-day PoC generation as a constraint-solving problem over oracle-related constraints.
\item At the technical level, we design several novel techniques to improve the effectiveness 
    of constraint solving in PoC generation, including oracle synthesis, oracle-guided generator evolution,
    and oracle-guided directed mutation.
\item At the practical level, we build the end-to-end system and conduct a comprehensive evaluation
    of 138 real-world vulnerabilities on the Magma benchmark, 
    demonstrating that DIG substantially outperforms state-of-the-art LLM-driven input generation systems, as well as directed and non-directed fuzzers, 
    in PoC generation. We further commit to open-sourcing our
    artifacts to facilitate reproducibility and future research.

\end{itemize}

\section{Motivating Examples}
\label{sec:motivating_examples}
In this section, we demonstrate DIG's effectiveness for PoC generation through
two case studies on one-day vulnerabilities in real-world programs. In
\S\ref{sec:motivating_example_pdf001}, we show how DIG's oracle-guided generator
evolution generates a PoC for CVE-2019-14494, where other state-of-the-art
techniques fail. In \S\ref{sec:motivating_example_tif008}, we show how DIG's
oracle-guided directed mutation generates a PoC for CVE-2015-8784, where
DIG's agent alone fails due to hidden program state.

\subsection{CVE-2019-14494}
\label{sec:motivating_example_pdf001}
This vulnerability is a floating-point underflow leading to division-by-zero 
in Poppler's PDF tiling pattern renderer (Figure~\ref{fig:pdf001_vuln_code} in Appendix). 
It occurs in \texttt{SplashOutputDev::tilingPatternFill()} when rendering 
specially crafted PDF files with tiling patterns.

To reach the vulnerable code, the input must satisfy standard PDF syntactic constraints:
valid PDF structure (e.g., a \texttt{\%PDF-1.4} header), correct object references, proper dictionary syntax,
and a well-formed TilingPattern object with all required fields (\texttt{/PatternType},
\texttt{/BBox}, \texttt{/XStep}, \texttt{/YStep}). These constraints are enforced by the
PDF parser and object construction. Any syntactic violation
causes parsing failure before reaching the vulnerable function.
The critical challenge lies in satisfying multiple semantic constraints at runtime. 
We highlight three essential ones:

\begin{itemize}[leftmargin=*, itemsep=3pt, topsep=3pt]
\item \textbf{Singularity check (C7).} Before invoking the vulnerable function \texttt{tilingPatternFill()},
Poppler validates that the Current Transformation Matrix (CTM) is not approximately singular
(Figure~\ref{fig:pdf001_singularity_check} in Appendix). In PDF, the CTM is represented as a six-parameter
affine transform \texttt{[a b c d e f]}, whose linear part is the $2 \times 2$ matrix
$\begin{pmatrix} a & c \\ b & d \end{pmatrix}$ with determinant $ad-bc$.
The check requires $|\det(\text{CTM})| \geq 10^{-6}$.
The challenge is to find a CTM whose linear part is large enough to pass this check,
yet still small enough to contribute to the downstream floating-point underflow.
For instance, a CTM with 
$a = 10^{-5}$, $b = 0$, $c = 0$, $d = 1$ yields $\det = 10^{-5} > 10^{-6}$, 
bypassing the check while enabling the subsequent floating-point underflow. In contrast, 
values like $a = 10^{-1}$ would pass the singularity check but fail to trigger underflow.

\item \textbf{XStep/YStep dependency (C8).} The Pattern's \texttt{/XStep} and \texttt{/YStep}
must equal the BBox dimensions in \texttt{SplashOutputDev::doTilingPatternFill()}. 
This cross-field dependency requires \texttt{xStep == bbox[2] - bbox[0]} and \texttt{yStep == bbox[3] - bbox[1]}.

\item \textbf{Floating-point underflow chain (C9--C10).} The most critical semantic constraint
involves a multi-step computation that must produce zero through floating-point underflow.
As shown in Figure~\ref{fig:pdf001_vuln_code}, the product of three values from different
PDF objects must underflow below the minimum positive subnormal value of IEEE 754
double precision ($\approx 4.94 \times 10^{-324}$ for C/C++ \texttt{double}).
This requires: (1) an extremely small tile width ($10^{-340}$) specified in the
Pattern's \texttt{/BBox} dictionary, (2) a small CTM scale factor ($10^{-5}$) from the 
transformation matrix (constrained by the singularity check), and (3) their product 
$10^{-5} \times 10^{-340} = 10^{-345} < 4.94 \times 10^{-324}$ underflowing to 
exactly zero. This constraint is particularly subtle because the \texttt{ceil()} function in 
\texttt{surface\_width = (int)ceil(fabs(kx))} rounds any non-zero value up to 
at least 1. For example, even a tiny value like $kx = 10^{-12}$ would result in 
\texttt{ceil}$(10^{-12}) = 1$, yielding \texttt{surface\_width = 1} and avoiding 
division-by-zero. Only when \texttt{kx} underflows to \emph{exactly} 0.0 will 
\texttt{surface\_width} become zero.

\end{itemize}

DIG first analyzes the vulnerability patch (Figure~\ref{fig:pdf001_patch} in Appendix) and identifies the key constraints
that are most relevant to the vulnerability. In this case, the patch adds a zero check before the division.
DIG then analyzes how this check prevents the
vulnerable behavior in the original code and infers that triggering the vulnerability
requires \texttt{surface\_width == 0 || surface\_height == 0}, which DIG uses as the oracle.

DIG generates a valid PoC (Figure~\ref{fig:pdf001_poc} in Appendix) through 12 rounds of generator evolution.
At each iteration, DIG conducts LLM-based code reasoning from the oracle to infer the constraints 
that govern how the variables in the oracle are computed and propagated. 
Figure~\ref{fig:dig_constraint_inference_pdf001} summarizes the constraints inferred by DIG across these iterations.

Early generators (cycles 1--9) focused on satisfying the syntactic constraints required to reach the oracle.
These included constructing a valid PDF structure and tiling pattern, populating the required Pattern fields, ensuring
that the Pattern is referenced by the page resources, and including the necessary fill operators in the content stream (i.e., C1--C5).
DIG also derived the semantic constraints governing the computation of the oracle, including 
the relation between \texttt{XStep}, \texttt{YStep}, and \texttt{BBox} (C8) and the numerical conditions 
required for the underflow chain (C9--C10).
A critical breakthrough came in cycle 9, when DIG inferred that \texttt{ceil(fabs(kx))} makes 
any positive value round up to at least 1, implying that the oracle can be satisfied only 
when \texttt{surface\_width} becomes exactly zero (C10).
This insight narrowed the remaining unresolved constraints to achieving exact zero through floating-point underflow. 
Guided by this constraint, cycle 10 introduced an extreme tile width ($10^{-340}$), and cycle 11 preserved this 
logic while further refining the CTM scale ($10^{-5}$) so that the remaining semantic constraints, 
including the non-singularity constraint (C7), were simultaneously satisfied.

This process illustrates how oracle-guided generator evolution mitigates the \emph{goal drift} problem
in agentic reasoning. 
By focusing each iteration on the currently unresolved constraints 
while preserving previously recovered generation logic, DIG incrementally builds the generator and gradually constructs the final PoC.
In our experiments, neither existing agentic PoC-generation systems, such as PBFuzz~\cite{zeng2025pbfuzz}, nor 
general-purpose coding agents, such as Cursor~\cite{anysphere2023cursor}, were able to satisfy the oracle and generate a valid PoC.
Moreover, state-of-the-art fuzzers, as well as
LLM-assisted input-generation approaches such as G$^2$Fuzz~\cite{zhang2025low}, LlamaFuzz~\cite{zhang2024llamafuzz}, and
SeedAIchemy~\cite{wen2025seedaichemy}, failed even to satisfy the syntactic constraints 
required to reach the vulnerable function.

\begin{figure}[htb!]
    \centering
    \adjustbox{width=\linewidth,center}{
    \begin{tcolorbox}[colback=gray!5, colframe=gray!40, title=DIG constraint inference for CVE-2019-14494, fonttitle=\bfseries]
    \footnotesize    
    \begin{itemize}[leftmargin=*, noitemsep]
        \item C1: $\textit{valid\_pdf\_structure}(\textit{file})$ (proper header and object structure)
        \item C2: $\textit{PatternType} = 1$ (TilingPattern)
        \item C3: $\exists \textit{BBox}, \textit{XStep}, \textit{YStep}, \textit{Matrix} \in \textit{Pattern}$ (required fields)
        \item C4: $\textit{Pattern} \in \textit{Page}.\textit{Resources}$ (resource reference)
        \item C5: $\textit{content\_stream} \supseteq \{\texttt{/Pattern cs}, \texttt{scn}, \texttt{re f}\}$ (fill operators)
        \item C6: $\forall i \in [0,5]: \textit{isfinite}(\textit{CTM}[i])$ (finite CTM values)
        \item C7: $|\det(\textit{CTM})| = |\textit{CTM}[0] \times \textit{CTM}[3] - \textit{CTM}[1] \times \textit{CTM}[2]| \geq 10^{-6}$
        \item C8: $\textit{XStep} = \textit{BBox}[2] - \textit{BBox}[0] \land \textit{YStep} = \textit{BBox}[3] - \textit{BBox}[1]$
        \item C9: $\textit{kx} \times \textit{tile\_width} \times (\textit{x1} - \textit{x0}) < 4.94 \times 10^{-324}$ (underflow)
        \item C10: $\textit{surface\_width} = \lceil |\textit{kx}| \rceil = 0$ (zero after scaling)
    \end{itemize}
    \end{tcolorbox}
    }
    \caption{Constraints inferred by DIG for CVE-2019-14494.}
    \label{fig:dig_constraint_inference_pdf001}
    \end{figure}

\subsection{CVE-2015-8784}
\label{sec:motivating_example_tif008}

\begin{figure}[htb!]
    \centering
    \adjustbox{width=\linewidth,center}{
    \begin{tcolorbox}[colback=gray!5, colframe=gray!40, title=Constraints required to trigger CVE-2015-8784, fonttitle=\bfseries]
    \footnotesize
    \begin{itemize}[leftmargin=*, noitemsep]
        \item C1: $\textit{valid\_tiff\_structure}(\textit{file})$ (valid magic number, IFD entries, and strip/tile data)
        \item C2: $\textit{Compression} = 32766$ (NeXT 2-bit RLE codec selected)
        \item C3: $\textit{BitsPerSample} = 2$ (only value accepted by \texttt{NeXTPreDecode})
        \item C4: $\textit{isTiled}(\textit{file})$ (\texttt{TileWidth} and \texttt{TileLength} tags present)
        \item C5: $\textit{td\_tilewidth} > \textit{td\_imagewidth}$
        (\texttt{scanline} is computed from \textit{td\_imagewidth} via \texttt{TIFFScanlineSize},
        while the decode loop uses \textit{td\_tilewidth} as \texttt{imagewidth};
        the mismatch allows \texttt{op\_offset} to reach \texttt{scanline}
        before \texttt{npixels} reaches \texttt{imagewidth})
    \end{itemize}

    \end{tcolorbox}
    }
    \caption{Constraints required to trigger CVE-2015-8784.}
    \label{fig:tif008_constraints}
\end{figure}

This vulnerability is a heap buffer overwrite in
\texttt{NeXTDecode} (Figure~\ref{fig:tif008_vuln_code} in Appendix).
To reach \texttt{NeXTDecode}, the input must satisfy syntactic
constraints C1--C3: a valid TIFF file using NeXT compression (tag 32766)
and \texttt{BitsPerSample}~=~2, the only value accepted by the
pre-decode validator \texttt{NeXTPreDecode}.
The overflow occurs because the inner \texttt{while} loop and the output
buffer use two different values for the row width: the loop iterates
until \texttt{npixels} reaches \texttt{imagewidth}, while the buffer
allocated for each output row is only \texttt{scanline} bytes.
When \texttt{imagewidth} exceeds what \texttt{scanline} can hold,
\texttt{SETPIXEL} writes past the end of the row into adjacent heap
memory.

As shown in Figure~\ref{fig:tif008_vuln_code}, \texttt{scanline} is
read from \texttt{tif->tif\_scanlinesize}, which is computed by
\texttt{TIFFScanlineSize64} (bottom of the figure) solely from
\texttt{td\_imagewidth}.
In the default \emph{strip} mode, \texttt{NeXTDecode} also sets the
local variable \texttt{imagewidth} from \texttt{td\_imagewidth}.
Since both values derive from the same field, \texttt{SETPIXEL} (which
packs four 2-bit pixels per byte, incrementing \texttt{op\_offset} every
fourth pixel) always stays within the buffer: by the time
\texttt{npixels} reaches \texttt{imagewidth}, \texttt{op\_offset} equals
\texttt{scanline} and the loop exits.
In \emph{tiled} mode (C4), the \texttt{if~(isTiled)} branch in the
figure overrides \texttt{imagewidth} with \texttt{td\_tilewidth}.
However, \texttt{scanline} is \emph{not} updated to match; it still
reflects \texttt{td\_imagewidth}.
When \texttt{td\_tilewidth} $>$ \texttt{td\_imagewidth} (C5), the loop
runs for more pixels than the buffer can hold.
For example, with \texttt{td\_imagewidth}~=~4 and
\texttt{td\_tilewidth}~=~16: \texttt{scanline}~=~1 byte (room for
4~pixels), but the loop runs for 16~pixels.
After the first 4~pixels, \texttt{op\_offset}~=~1~$\geq$~%
\texttt{scanline}~=~1
while \texttt{npixels}~=~4~$<$~\texttt{imagewidth}~=~16, and
\texttt{SETPIXEL} begins writing out of bounds.

In our experiments, DIG successfully analyzed the oracle derived from the patch,
but its generator evolution failed to generate a valid PoC. We found that the agent failed to
infer the hidden state relationship between two variables: how
\texttt{scanline} and \texttt{imagewidth} are computed, and how tiled mode
decouples them. In strip mode, both values are derived from \texttt{td\_imagewidth}; in tiled mode, \texttt{imagewidth} is overwritten with
\texttt{td\_tilewidth}, whereas \texttt{scanline} still reflects
\texttt{td\_imagewidth}. 

In contrast, oracle-guided directed mutation does not require understanding the hidden relationship. Instead, it observes
how each mutation changes the oracle branch distance. Adjusting the RLE stream can
initially increase the number of decoded pixels, reducing the distance to
\texttt{op\_offset >= scanline}. However, in strip mode, \texttt{npixels} is
bounded by \texttt{imagewidth}, so \texttt{op\_offset} eventually stops growing
and the branch distance stops decreasing. The fuzzer can then explore other input
regions that affect the oracle. A byte-level mutation that happens to create a valid tiled-mode configuration
(e.g., by setting the \texttt{TileWidth} and \texttt{TileLength} entries) can
make \texttt{isTiled(tif)} true, causing \texttt{imagewidth} to be overwritten
with \texttt{td\_tilewidth}.
This immediately changes the branch distance. A larger
\texttt{td\_tilewidth} raises the upper bound on \texttt{op\_offset}, allowing
the distance to \texttt{op\_offset >= scanline} to decrease again. Therefore, 
further insertions into the RLE stream become effective again,
increasing \texttt{op\_offset} until the out-of-bounds write is triggered. 
In our experiments, DIG's low-level directed mutation triggered the vulnerability
13.1$\times$ faster than state-of-the-art mutation-enhanced fuzzers such as AFL++~\cite{fioraldi2020afl++} and FOX~\cite{she2024fox}.

\FloatBarrier

\section{DIG Design}
In this section, we first describe DIG's overall architecture and key
components (\S\ref{sec:dig_workflow}). We then define the oracle and explain
how it is synthesized (\S\ref{sec:oracle_synthesis}). Next, we detail DIG's
oracle-guided generator evolution (\S\ref{sec:oracle-guided-generator-evolution}), and finally its 
oracle-guided directed mutation (\S\ref{sec:oracle-guided-directed-mutation}).

\begingroup

\setlength{\belowcaptionskip}{6pt}
\begin{figure}[t]
    \centering
    \includegraphics[width=\linewidth]{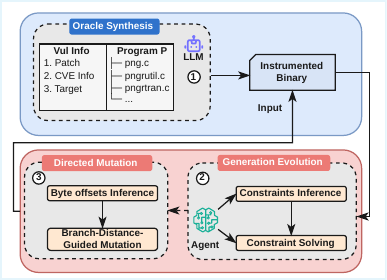}
    \caption{Overview of DIG's workflow for PoC generation.}
    \label{fig:dig_workflow}
\end{figure}
\endgroup

\subsection{DIG Workflow}
\label{sec:dig_workflow}
Figure~\ref{fig:dig_workflow} illustrates the overall workflow of DIG.
\circled{1} Given patch information, the CVE description, and the program source code, 
DIG synthesizes an oracle and instruments it into the target program. 
It also enables sanitizer instrumentation to detect whether candidate inputs actually trigger the vulnerability.
\circled{2} DIG's agent infers constraints from the oracle by invoking a suite of program analysis tools, including static,
dynamic, and source-code analyzers. Based on the inferred constraints, DIG constructs and refines a generator
that preserves previously satisfied constraints while learning to satisfy unresolved ones.
\circled{3} When high-level generator evolution fails to satisfy the oracle,
DIG performs low-level directed mutation on inputs generated by the generator. In this step, DIG performs black-box taint analysis
to identify the input bytes that directly influence the oracle. Then, DIG mutates these bytes based on the branch distance feedback.
Step \circled{2} is repeated until the oracle is satisfied or the generator-evolution budget is reached.
Step \circled{3} is repeated until the oracle is satisfied or a
timeout is reached.

\subsection{Oracle Synthesis}
\label{sec:oracle_synthesis}
\begin{definition}[Oracle]
  Let $P$ be a program under test, $\mathcal{I}$ its input space, and $S$ the set of
  runtime program states. We write $\langle P, i \rangle \to^* s$ to denote that
  executing $P$ on input $i \in \mathcal{I}$ can reach state $s \in S$.
  Let $S_v \subseteq S$ be the set of vulnerable states that characterize
  the triggering of a vulnerability $v$.

  An \textbf{\emph{oracle}} is a patch-derived boolean predicate
  $O : S \to \{\mathsf{True}, \mathsf{False}\}$ such that any execution that
  reaches $S_v$ must pass through a state satisfying $O$:
  \[
  \forall i \in \mathcal{I},\;
  \bigl(\langle P,i\rangle \to^* S_v\bigr)
  \Rightarrow
  \bigl(\exists s \in S.\; \langle P,i\rangle \to^* s
  \land O(s) = \mathsf{True}\bigr).
\]
\end{definition}
An important aspect of this definition is that the oracle is patch-derived.
This distinguishes DIG’s oracle from arbitrary necessary preconditions, such as dominator branches along the execution path.
In contrast, the oracle is derived from the patch and therefore encodes a specific precondition for vulnerability triggering.
This follows from the assumption that the patch correctly and completely fixes the vulnerability. 
A security patch typically introduces checks or enforces conditions that were missing in the vulnerable version.
Therefore, any execution that triggers the vulnerability must violate these "patch-enforced" conditions. 
If a triggering execution did not violate them, the patch would fail to prevent that execution, 
contradicting the assumption that the patch correctly and completely fixes the vulnerability.

DIG’s oracle should not be confused with the check used by sanitizers.
A sanitizer check is a detection mechanism that monitors whether a final execution outcome, 
such as memory corruption or a crash, has occurred. In contrast, DIG’s oracle is used to guide the search toward PoC generation.
It captures patch-derived vulnerability conditions that inputs must satisfy before the final bug manifestation can occur. 
A natural question, then, is whether a crash condition can serve as such a search oracle.
In principle, crash conditions are often unavailable.
Many vulnerabilities do not come with bug reports. For example, in our
study of 20 OpenSSL CVEs, none were accompanied by a bug report. In
contrast, patch information is consistently available.

DIG synthesizes an oracle through LLM-based analysis of vulnerability
artifacts: the vulnerable source code, bug description, and security patch.
Figure~\ref{fig:oracle_prompt} shows a simplified version of the prompt that guides this
process. The prompt first defines an oracle as a patch-derived necessary
precondition for vulnerability triggering. It then guides the LLM to synthesize 
the oracle in three steps. First, it
identifies the constraints in the patch that are most relevant to the vulnerability, such as a
newly added check, a corrected variable relation, or an enforced state
condition. Second, it analyzes how these constraints prevent the original
vulnerable behavior in the context of the source code and vulnerability
description. Third, it infers what must hold in the vulnerable version for
these constraints to be violated, which DIG uses as the oracle.  

\begin{figure}[H]
  \begin{tcolorbox}[colback=blue!5, colframe=blue!40, title=Oracle Synthesis Prompt, fonttitle=\bfseries]
  \footnotesize
  \textbf{<role>} 
  You are an expert security researcher specializing in patch analysis and oracle synthesis.
  An \textbf{oracle} is a patch-derived boolean predicate over program states that captures
  a necessary precondition for vulnerability triggering.

  \textbf{<mission>}
  Your mission is to synthesize such an oracle by analyzing the security patch, vulnerable
  source code, and bug description. First, identify the constraints in the patch that are most relevant to
  the vulnerability, such as a newly added check, a corrected variable relation, or an enforced
  state condition. Second, analyze how these constraints prevent the vulnerable behavior 
  in the source code and vulnerability description. Third, infer what must hold in the source code for these constraints
  to be violated, and express these violations as an oracle.

  \textbf{<input>}
  \begin{itemize}[leftmargin=*, noitemsep, topsep=2pt]
      \item \textbf{Security Patch (Diff):} \texttt{<PATCH>}
      \item \textbf{Source Code to the PATCH:} \texttt{<SOURCE\_CODE>}
      \item \textbf{Vulnerability Description:} \texttt{<DESCRIPTION>}
  \end{itemize}
  
  \textbf{<output>}
  Return a C/C++ code snippet that specifies both where the oracle should be
  inserted in the vulnerable source code and what predicate should be checked.
  The predicate should use in-scope program variables at that location. 

  \end{tcolorbox}
  \caption{Simplified prompt for oracle synthesis.}
  \label{fig:oracle_prompt}
\end{figure}

\subsection{Oracle-guided Generator Evolution}
\label{sec:oracle-guided-generator-evolution}
Our oracle-guided generator evolution consists of two phases: constraint inference 
and constraint solving. Algorithm~\ref{alg:dig_main} illustrates the overall process.
For constraint inference (lines~6--7), we use an ACTION-OBSERVATION agentic workflow~\cite{yao2022react}
to infer oracle-related constraints
(more details in \S\ref{sec:oracle-guided-constraint-inference}).
For constraint solving (lines~8--14), we define the effectiveness of generators based on their
constraint satisfaction sets, and select the most effective generator to solve the constraints
(more details in \S\ref{sec:oracle-guided-constraint-solving}).

Both phases share a common agentic tool set that the agent invokes during reasoning:

\noindent\textbf{Static Analysis Tools} reason about interprocedural control flow:
\begin{itemize}[noitemsep, topsep=2pt]
  \item \texttt{get\_call\_path(\textit{func})} --- returns the call path from \texttt{main} to the target.
  \item \texttt{get\_callers(\textit{func}, \textit{depth})} --- returns callers up to a given depth.
\end{itemize}

\noindent\textbf{Source Code Inspection Tools} examine oracle-related variable definitions:
\begin{itemize}[noitemsep, topsep=2pt]
  \item \texttt{get\_source\_code(\textit{func})} --- retrieves function source.
  \item \texttt{grep\_code(\textit{pattern})} --- locates assignment sites of oracle-related
        variables across the codebase.
\end{itemize}

\noindent\textbf{Dynamic Analysis Tools} compare runtime behavior against inferred constraints:
\begin{itemize}[noitemsep, topsep=2pt]
  \item \texttt{get\_execution\_trace()} --- records execution trace on a candidate input.
  \item \texttt{analyze\_condition()}  --- inspects constraint satisfaction on the execution trace.
\end{itemize}

\subsubsection{Oracle-guided Constraint Inference}
\label{sec:oracle-guided-constraint-inference}
As we discussed in \S\ref{sec:intro}, the key challenge of PoC generation is to identify which constraints are necessary to 
trigger the vulnerability and solve them effectively.
Through oracle synthesis (\S\ref{sec:oracle_synthesis}), DIG identifies a necessary precondition derived from the patch for vulnerability triggering. 
However, this oracle does not fully specify all the necessary constraints that must be satisfied to 
reach and satisfy the oracle. Therefore, DIG performs constraint inference to recover these constraints and uses them to guide generator evolution.

Specifically, DIG performs constraint inference through an ACTION-OBSERVATION agentic workflow. As shown in Algorithm~\ref{alg:dig_main}, the function 
\texttt{UpdConstr} is called at line 6 in the ACTION phase. The agent inspects the source code (via \texttt{get\_source\_code} and \texttt{grep\_code}) 
and invokes static analysis tools (via \texttt{get\_call\_path} and \texttt{get\_callers}) to reason about data flow and control flow,
inferring additional constraints necessary for satisfying the oracle.
If a newly inferred constraint is not already included in the oracle-related constraint set, the agent adds it to the set.
The function \texttt{UpdUnsatConstr} is called at line 7 in the OBSERVATION phase. During this phase, the agent executes the most 
effective generator so far with dynamic analysis tools such as \texttt{get\_execution\_trace} 
and \texttt{analyze\_condition} to inspect which constraints are satisfied and which are not. 
Then, the unsatisfied constraints are added to the unsatisfied constraint set, which is used to guide 
the generator refinement at line 13 to generate more effective generators to solve the unsatisfied constraints.

\begin{algorithm}[t]
  \caption{Oracle-Guided Generator Evolution}
  \label{alg:dig_main}
  \small
  \begin{algorithmic}[1]
    \Require Oracle $O$, instrumented program $P$, source code $\mathit{src}(P)$, budget $B$
    \State $\Sigma_O \gets O$ \Comment{oracle-related constraints}
    \State $\Sigma_U \gets \emptyset$ \Comment{unsatisfied constraints}
    \State $G \gets \emptyset$ \Comment{set of refined generators}
    \State $g^* \gets \bot$ \Comment{most effective generator}
    \While{$B > 0$}
      \Statex \hspace{2.2em}\textit{// Phase 1: Constraint Inference}
      \State $\Sigma_O \gets \Call{UpdConstr}{\Sigma_O, \mathit{src}(P)}$ \Comment{ACTION phase}
      \State $\Sigma_U \gets \Call{UpdUnsatConstr}{\Sigma_O, P, g^*}$ \Comment{OBSERVATION phase}
      \Statex
      \Statex \hspace{2.2em}\textit{// Phase 2: Constraint Solving}
      \State $g^* \gets \operatorname*{arg\,max}_{g \in G} \text{Eff}(g)$ \Comment{select most effective generator}
      \State $i \gets g^*()$ \Comment{generate input from best generator}
        \If{$\Call{Execute}{i, P}$ triggers crash}
          \State \Return $i$
        \Else
          \State $g \gets \Call{RefineGenerator}{\Sigma_U, g^*}$
          \State $G \gets G \cup \{g\}$
        \EndIf
      \State $B \gets B - 1$
    \EndWhile
    \State \Return $g^*$
  \end{algorithmic}
\end{algorithm}

\subsubsection{Oracle-guided Constraint Solving}
\label{sec:oracle-guided-constraint-solving}
To guide the generator evolution, we define the effectiveness 
of generators based on their constraint satisfaction sets.
Given an oracle $O$, let $\Sigma_O$ denote the set of
oracle-related constraints inferred from $O$.

\begin{definition}[Constraint Satisfaction Set]
  For a generator $g$, its constraint satisfaction set is:
  $$\text{CS}(g,\Sigma_O) = \{c \in \Sigma_O \mid \text{inputs from } g \text{ satisfy } c\}$$
\end{definition}

\begin{definition}[Effectiveness of Generators]
  Generator $g^*$ is more effective than $g$ if 
  $\text{CS}(g, \Sigma_O) \subset \text{CS}(g^*,\Sigma_O)$; in other words, $g$ is dominated by $g^*$. We denote this as:
  $$\text{Eff}(g) < \text{Eff}(g^*)$$
  where $\text{Eff}(\cdot)$ represents the effectiveness of a generator. 
\end{definition}

\begin{figure}[htbp]
  \centering
  \includegraphics[width=0.75\columnwidth]{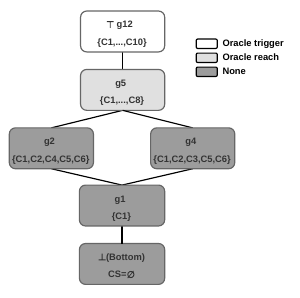}
  \caption{Generator lattice for CVE-2019-14494, showing the evolution of 
  generators through constraint satisfaction.}
  \label{fig:generator_lattice}
\end{figure}

The generator evolution process satisfies oracle-related constraints through a lattice structure 
defined by the effectiveness relation. The subset-based comparison of constraint satisfaction sets 
creates a partial order over generators. The bottom element $\bot$ represents the least 
effective generator that satisfies no constraints, while the top element $\top$ 
represents the most effective generator where $\text{CS}(g^*,\Sigma_O) = \Sigma_O$. 
Generators between these two are 
ordered by the subset relation of their constraint satisfaction sets: a proper 
superset indicates strictly greater effectiveness, while two generators are 
incomparable when they satisfy different constraint subsets such that neither 
subset fully encompasses the other.

At line~8 in Algorithm~\ref{alg:dig_main}, DIG selects the most effective generator $g^*$
by maximizing $\text{Eff}(g)$ over the current generator set $G$.
The effectiveness of each generator is measured via \texttt{analyze\_condition()} (via GDB),
which inspects the concrete variable values at each constraint site to determine
which constraints in $\Sigma_O$ are satisfied by inputs from $g$.
Because the effectiveness relation is a partial order, two generators may be
\emph{incomparable}: neither dominates the other when they satisfy different but overlapping
constraint subsets.
In such cases, DIG randomly selects among the non-dominated generators in $G$.
At line~13, the function \texttt{RefineGenerator} is called to refine the selected generator $g^*$.
This function invokes source code inspection tools (\texttt{get\_source\_code}, \texttt{grep\_code})
to locate where unsatisfied constraints originate in the source code,
then calls an LLM to update the generator to satisfy those constraints.

Figure~\ref{fig:generator_lattice} visualizes the lattice structure for 
CVE-2019-14494 (\S\ref{sec:motivating_example_pdf001}), 
where the superscript in $g^{x}$ denotes the iteration number.
Generators are color-coded by their oracle-related status.
Generators $g^{2}$ and $g^{4}$ are 
incomparable as they satisfy different constraint subsets. The evolution 
progresses from $\bot$ (0 constraints) 
through $g^{5}$ (8 constraints, first to reach the oracle) to $g^{12}$ 
(satisfying all constraints in $\Sigma_O$).

\subsection{Oracle-guided Directed Mutation}
\label{sec:oracle-guided-directed-mutation}
When the generator evolution fails to satisfy the oracle or the budget is reached, DIG performs directed mutation to guide the search toward the oracle.
Algorithm~\ref{alg:oracle_guided_mutation} presents our oracle-guided directed 
mutation strategy.
First, in the byte-offset influence phase, also known as black-box taint 
analysis~\cite{gan2020greyone, aschermann2019redqueen} (Lines 2--9), the algorithm 
identifies which byte offsets in the seed input $s$ influence the oracle's 
branch distance $\delta_O(s)$. For each byte offset $i$, \textsc{MutateByte}
mutates the byte and checks whether the branch distance changes. 
Only offsets that affect branch distance are considered influential.
Second, in the mutation-with-gradient-feedback phase (Lines 10--13), for each 
influential byte offset, \textsc{ComputeGradient} 
computes a gradient $g_i$ that estimates how sensitive the oracle's branch distance
function is to changes in the byte. Specifically, we approximate 
the gradient as $g_i \approx \frac{\delta_O(s[i] + \epsilon) - \delta_O(s[i])}{\epsilon}$, where 
$\epsilon$ is a small perturbation (e.g., $\epsilon = 1$). This gradient provides 
an estimated direction of change to steer input bytes toward satisfying the oracle.
The algorithm then mutates input bytes in the direction that 
reduces the branch distance, which provides gradient feedback for 
mutations.

Compared to random mutation, oracle-guided directed mutation offers two key 
advantages. First, it focuses mutations on input regions directly influencing 
the oracle. For CVE-2015-8784 (\S\ref{sec:motivating_example_tif008}), DIG mutates the bytes that influence the oracle \texttt{op\_offset >= scanline}
rather than performing random mutations that cause parsing failures and exit early. 
Second, the oracle provides branch distance as gradient feedback~\cite{chen2018angora, she2024fox}.
Given \texttt{op\_offset = 100, scanline = 500}, the branch distance of 400 indicates 
adjustment magnitude; the \texttt{>=} operator indicates direction 
(\texttt{op\_offset} must increase). Random mutation lacks both: it cannot 
distinguish \texttt{op\_offset = 105} from \texttt{op\_offset = 100} (both 
fail but one is closer to satisfying the oracle), and 
half its mutations move in the wrong direction.

\begin{algorithm}[t]
  \caption{Oracle-guided Directed Mutation}\label{alg:oracle_guided_mutation}
  \begin{algorithmic}[1]
  \Require Seed $s$ reaching oracle $O$, branch distance function $\delta_O$ for $O$
  
  \State $I \gets \emptyset$ 
      \Comment{Byte offsets in $s$ that influence $\delta_O(s)$} 
  \For{each byte offset $i$ in $s$}
      \State $d \gets \delta_O(s)$ \Comment{Branch distance before mutation}
      \State $s[i] \gets$ \Call{MutateByte}{$s[i]$}
      \State $d' \gets \delta_O(s)$
      \If{$d \neq d'$}
          \State $I \gets I \cup \{i\}$
      \EndIf
  \EndFor
  
  \For{each $i \in I$}
      \State $g_i \gets$ \Call{ComputeGradient}{$s$, $i$, $\delta_O$}
      \State $s[i] \gets$ \Call{MutateByte}{$s[i]$, $s[i] - g_i$}
  \EndFor
  
  \State \Return $s$
  
  \end{algorithmic}
  \end{algorithm}

\section{Implementation}
\label{sec:implementation}

We implement three main components in DIG: oracle synthesis, oracle-guided generator evolution, and oracle-guided directed mutation. 
Oracle synthesis is implemented in Python with 1,456 LoC. Oracle-guided generator evolution is implemented in Python and C++ with 14,163 LoC.
Directed mutation is implemented in C/C++ based on AFL~\cite{zalewski_afl}, using LLVM 14 with 1,212 LoC.

As part of generator evolution, DIG includes an agentic tool set to
support constraint inference and input generation. This tool set consists of three 
types of tools: static analysis, dynamic analysis, and source-code inspection.
The static analysis tools, \texttt{get\_call\_path} and \texttt{get\_callers}, are implemented 
in C++ using LLVM 14, with 1,671 LoC. 
To implement these tools, we build a call graph with indirect-call support based on type analysis.
The dynamic analysis tool \texttt{get\_execution\_trace} is implemented in C++ with 278 LoC;
it instruments target programs at compile time using GCC/Clang's \texttt{-finstrument-functions} flag,
which inserts \texttt{\_\_cyg\_profile\_func\_enter} callbacks at function entries to record call
sequences and invocation counts. Another dynamic analysis tool, \texttt{analyze\_condition},
is a lightweight wrapper around GDB that evaluates branch conditions at runtime. For source-code inspection tools,
\texttt{get\_source\_code} is implemented with tree-sitter and consists of 851 LoC, while \texttt{grep\_code} is a wrapper around \texttt{grep}.

\section{Evaluation}
In this section, we evaluate DIG to answer the following research questions:

\noindent\textbf{RQ1 (Oracle Synthesis Correctness):} 
How accurately can DIG synthesize vulnerability oracles compared to human experts?
\par\noindent\textbf{RQ2 (PoC Generation Effectiveness):} 
How effective is DIG at generating PoCs for known CVEs compared to 
state-of-the-art approaches?
\par\noindent\textbf{RQ3 (In-depth Analysis):} 
What insights can be gained from a detailed analysis of DIG in PoC generation?
\par\noindent\textbf{RQ4 (Ablation Study):} 
How do individual components (oracle-guided generator evolution, and oracle-guided directed mutation) contribute 
to DIG's overall effectiveness in PoC generation?
\par\noindent\textbf{RQ5 (New Vulnerability Finding):} 
Can DIG discover previously unknown vulnerabilities in real-world programs?

\subsection{Evaluation Setup}
We follow established best practices for rigorous fuzzing evaluation~\cite{klees2018evaluating, schloegel2024sok} 
and describe our experimental setup below. 

\subsubsection{Dataset}
We use the Magma benchmark~\cite{hazimeh2020magma} for four reasons:
(1) Real-world vulnerabilities: Magma consists of real-world vulnerabilities, 
covering 138 CVEs across 9 widely used open-source projects.
(2) Generality and scalability: the benchmark includes software projects with diverse input formats 
and a wide range of codebase sizes (see Table~\ref{tab:magma_target_size}).
(3) Standard and comparable benchmark: Magma has been widely adopted in prior fuzzing 
research~\cite{shah2022mc2, bao2025alarms, huang2024titan, she2024fox, zeng2025pbfuzz}, 
making it a commonly used standard for evaluating PoC generation capabilities.
(4) Unsaturated and challenging: according to the Magma documentation, 
only a subset of its vulnerabilities (e.g., 42 out of 138 CVEs) have been successfully triggered by existing fuzzers, 
indicating that the benchmark remains far from saturated and still leaves substantial room for improvement.

\subsubsection{Baselines and Setup}
We select two agentic systems and ten fuzzers as baselines (see Table~\ref{tab:fuzzing_baselines}). These baselines cover five categories:
(i) \textbf{C0}: Agentic input generation (PBFuzz~\cite{zeng2025pbfuzz} and Cursor~\cite{anysphere2023cursor});
(ii) \textbf{C1}: baseline coverage-guided fuzzing (AFL~\cite{zalewski_afl} and AFL++~\cite{fioraldi2020afl++});
(iii) \textbf{C2}: directed fuzzing employing various techniques (AFLGo~\cite{bohme2017directed}, SelectFuzz~\cite{luo2023selectfuzz}, Titan~\cite{huang2024titan});
(iv) \textbf{C3}: coverage-guided fuzzing with enhanced mutation (CmpLog~\cite{aschermann2019redqueen} and FOX~\cite{she2024fox}); and
(v) \textbf{C4}: LLM-assisted input generation added on top of baseline
coverage-guided fuzzing (G$^2$Fuzz~\cite{zhang2025low}, LLAMAFUZZ~\cite{zhang2024llamafuzz}, SeedAIchemy~\cite{wen2025seedaichemy}).
For all tools, we used the most recent stable release available at the
time of our study. For tools whose implementations are not publicly available at the time of writing,
such as SeedAIchemy, or whose deployment is difficult, such as LlamaFuzz, we directly used 
the results reported in the original papers because they also use the same benchmark for evaluation.

For fuzzers built on AFL-based engines, including AFL++, AFLGo, SelectFuzz, Titan, CmpLog, 
FOX, and G$^2$Fuzz, we configured the fuzzing engine in non-deterministic mode. 
Each non-LLM fuzzer (C1, C2, C3) was run for 24 hours per target and repeated 10 times
using Magma’s initial seed corpus. In contrast, LLM-based approaches (C0, C4),
including G$^2$Fuzz, PBFuzz, and Cursor, were run once due to cost constraints
and did not require Magma’s initial seed corpus. DIG was also run once per target
due to cost constraints, consistent with the settings used for G$^2$Fuzz, PBFuzz, and Cursor.
For PBFuzz and Cursor (configured with the prompt template in Figure~\ref{fig:cursor_initial_system_prompt}), 
we directly used the results reported by the PBFuzz authors.
To ensure a fair comparison between LLM-assisted and agentic input-generation
approaches, all such approaches used the same default model, Claude Sonnet 4.5.

In addition, all experiments were conducted on a server with
two AMD EPYC 9354 CPUs, 64 cores in total, and 256 GB RAM. 
For consistency, each instance was executed in a Docker container
limited to one CPU core and 4 GB RAM.

\subsubsection{Evaluation Metrics}

We evaluate PoC generation using two primary metrics. 
First, \textbf{CVE coverage} measures the number of distinct Magma CVEs successfully 
triggered. For non-LLM fuzzers (C1, C2, C3), CVE coverage is computed as the union of CVEs 
triggered across all 10 runs. For LLM-based approaches (C0, C4), i.e., G$^2$Fuzz, PBFuzz, Cursor, and DIG, 
which are executed once due to cost constraints, CVE coverage is computed as the 
total number of CVEs triggered in the single 24-hour run.

Second, \textbf{Time-to-Event (TTE)} measures the elapsed time from the start of 
fuzzing until the first successful PoC is generated. To account for runs that do 
not generate a PoC within the 24-hour time budget, we treat such runs as 
right-censored and estimate the survival function using the Kaplan--Meier 
estimator~\cite{kaplan1958nonparametric}. We report the restricted mean TTE (RMST), 
computed as the area under the survival curve up to the 24-hour cutoff.
For CVEs that can be triggered from multiple fuzzing harnesses, we report the minimum TTE.

\subsection{Oracle Synthesis Correctness}
Before assessing oracle correctness, we first establish a \emph{ground-truth oracle} for evaluation.
The Magma benchmark provides a per-vulnerability canary function (\texttt{MAGMA\_LOG}),
which implements a set of predicates intended to characterize necessary conditions for vulnerability triggering.
We treat this canary function as the \emph{reference oracle} and additionally perform
manual inspection to validate its intended semantics.
When discrepancies are observed, human judgment is treated as the final arbiter,
resulting in a ground-truth oracle.

Using this ground-truth oracle, we assess the correctness of DIG’s synthesized oracles
at three levels of increasing permissiveness.
At \textbf{L1 (Syntactic Equivalence)}, a synthesized oracle is considered correct if it is
syntactically identical to the ground-truth oracle, i.e., the two oracles contain exactly
the same predicates with identical structure.
At \textbf{L2 (Semantic Equivalence)}, a synthesized oracle is considered correct if it is
semantically equivalent to the ground-truth oracle, meaning that both oracles characterize
the same conditions, even if their predicate formulations differ syntactically.
At \textbf{L3 (Sufficient-Condition Correctness)}, a synthesized oracle is considered correct 
if it captures a sufficient condition for the ground-truth oracle. 
In other words, whenever the synthesized oracle evaluates to true,
the ground-truth oracle would also evaluate to true, even though
the synthesized oracle may be more restrictive and may not cover all cases captured by the ground-truth oracle.
In addition to these correctness levels, 
we distinguish two categories of errors encountered during evaluation: a \textbf{Synthesis Error}, 
where the synthesized oracle is incorrect with respect to the ground-truth oracle, and a \textbf{Reference Error}, 
where the reference oracle itself is incorrect with respect to the ground-truth oracle (e.g., the Magma canary).

\subsubsection{Synthesized-oracle Correctness.}
The Magma benchmark contains a total of 138 CVEs.
Three CVEs (\texttt{PNG004}, \texttt{PNG005}, and \texttt{XML015}) were excluded due to missing
or incorrect CVE annotations in the benchmark.
This leaves 135 synthesized oracles for evaluation against the ground-truth oracle.
Among the 135 evaluated cases, 122 synthesized oracles are correct under the L1-L3 criteria,
corresponding to a success rate of \textbf{90.37\%}.
The remaining 13 cases are classified as Synthesis Errors.
In addition, we identified 4 cases where the Magma reference oracle itself is incorrect,
which are categorized as Reference Errors. 
Table~\ref{tab:oracle_correctness} summarizes the results. 
Appendix~\ref{sec:oracle_case_studies} provides case studies for all categories discussed above: L1, L2, L3, Synthesis Error, and Reference Error.

\begin{tcolorbox}[colback=gray!10, colframe=black, boxrule=0.5pt, arc=2pt, left=6pt, right=6pt, top=6pt, bottom=6pt]
\textbf{RQ1 Takeaway:} Given patch information, DIG achieves a \textbf{90.37\%} 
success rate in synthesizing correct oracles across 135 CVEs. 
This suggests that LLM-guided reasoning is a promising approach for automating oracle synthesis at scale.
\end{tcolorbox}

\begin{table}[!t]
    \centering
    \caption{Oracle synthesis correctness breakdown across the Magma benchmark.}
    \label{tab:oracle_correctness}
    {\scriptsize
    \setlength{\tabcolsep}{3pt}
    \begin{tabularx}{\columnwidth}{X X X X X X X}
    \toprule
    \textbf{Project} &
    \textbf{\shortstack[c]{CVEs}} &
    \textbf{L1} &
    \textbf{L2} &
    \textbf{L3} &
    \textbf{\shortstack[c]{Syn.\\Err.}} &
    \textbf{\shortstack[c]{Ref.\\Err.}} \\
    \midrule
    libpng & 5 & 1 & 4 & 0 & 0 & 0 \\
    libsndfile & 18 & 12 & 3 & 0 & 3 & 2 \\
    libtiff & 14 & 8 & 4 & 0 & 2 & 1 \\
    libxml2 & 16 & 10 & 4 & 0 & 2 & 0 \\
    Lua & 4 & 1 & 2 & 1 & 0 & 0 \\
    poppler & 22 & 17 & 1 & 3 & 1 & 0 \\
    sqlite3 & 20 & 14 & 3 & 2 & 1 & 0 \\
    OpenSSL & 20 & 10 & 9 & 0 & 1 & 0 \\
    PHP & 16 & 8 & 4 & 1 & 3 & 1 \\
    \bottomrule
    \end{tabularx}
    }
\end{table}

\subsection{PoC Generation Effectiveness}

\subsubsection{CVE Coverage}
\begin{figure}[!htb]
    \centering
    \includegraphics[width=\columnwidth]{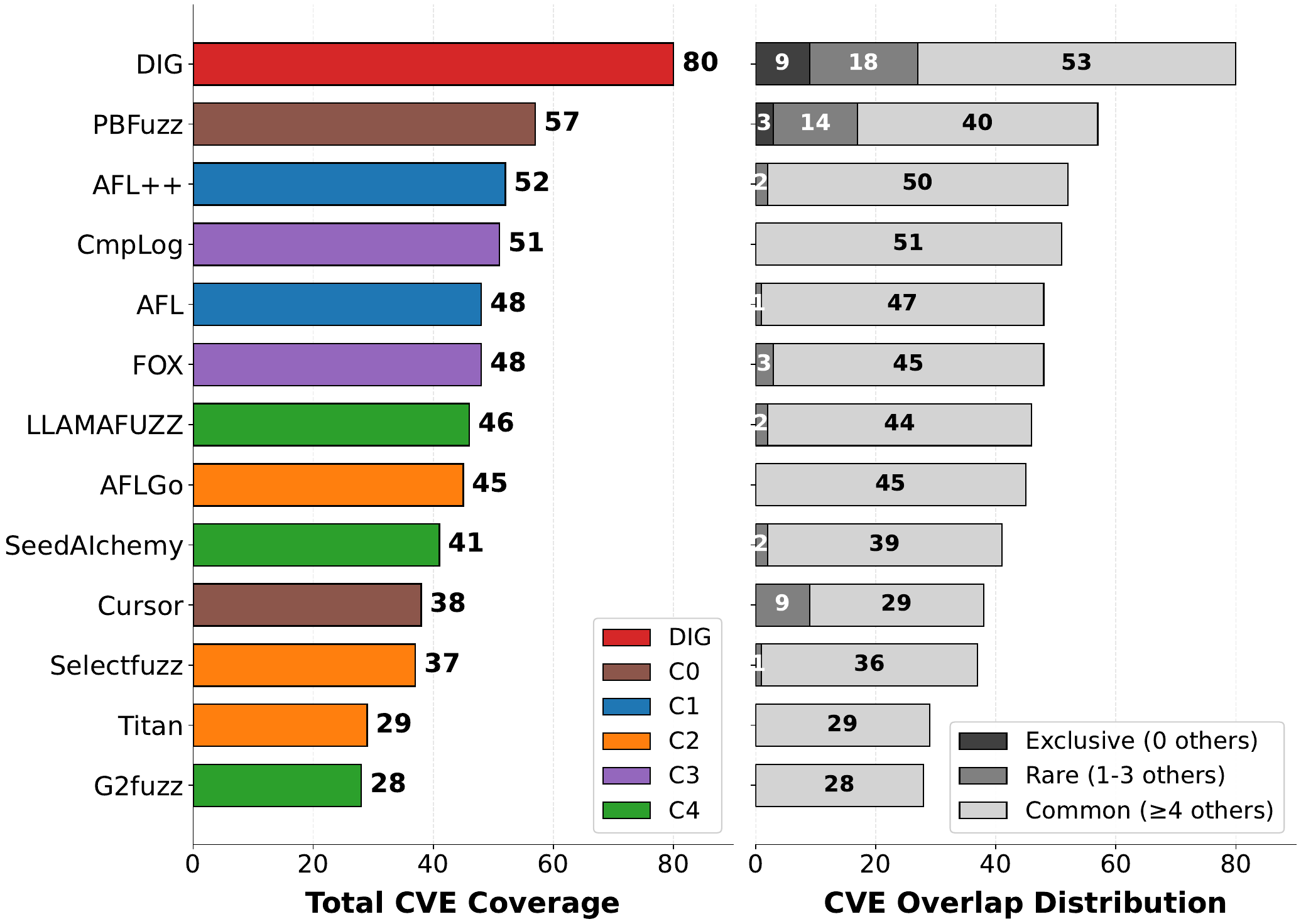}
    \caption{Left: CVE coverage per fuzzer. 
    Right: CVE coverage overlap across fuzzers.}
    \label{fig:coverage_overlap}
\end{figure}

To comprehensively evaluate DIG's effectiveness, we compare its CVE 
coverage against two agentic systems and 10 state-of-the-art fuzzers across all 138 CVEs in the Magma
benchmark detailed in Table~\ref{tab:trigger}.
Figure~\ref{fig:coverage_overlap} (left) shows that
DIG successfully triggers 80 out of 138 CVEs (58.0\%), substantially outperforming 
all tools. Agentic systems (C0) achieve 57 CVEs for PBFuzz (41.3\%) and 38 CVEs for Cursor (27.5\%).
Baseline fuzzers (C1) achieve 48 CVEs for AFL (34.8\%) 
and 52 CVEs for AFL++ (37.7\%). Directed fuzzers (C2) range from 28-45 CVEs: 
AFLGo leads with 45 CVEs (32.6\%), followed by SelectFuzz at 37 CVEs (26.8\%), and Titan at 29 CVEs (21.0\%). Mutation-enhanced 
fuzzers (C3) achieve 51 CVEs for CmpLog (37.0\%) and 48 CVEs for FOX (34.8\%). 
LLM-assisted input generation tools (C4) range from 28-46 CVEs: LLAMAFUZZ achieves 46 CVEs 
(33.3\%), SeedAIchemy 41 CVEs (29.7\%), and G$^2$Fuzz 28 CVEs (20.3\%).

To understand the unique CVEs found by different fuzzers,
we analyze the overlap distribution of the 84 CVEs across all 13 tools (including DIG).
Figure~\ref{fig:coverage_overlap} (right) 
shows, for each tool, how many of its CVEs are also found by other tools.
We define three overlap categories: Exclusive (0 others)---CVEs found 
by only one tool; Rare (1-3 others)---CVEs found by 2-4 tools 
total; Common ($\geq$4 others)---CVEs found by 5+ tools.
DIG has 9 exclusive and 18 rare CVEs out of its 80 total, meaning over one-third of 
DIG's coverage comes from vulnerabilities that few or no other tools can trigger. 
Among agentic systems (C0), PBFuzz has 3 exclusive and 14 rare CVEs and Cursor has 
no exclusive CVEs but contributes 9 rare ones, demonstrating the potential of 
agentic systems to trigger vulnerabilities that traditional fuzzers cannot. 
In contrast, all C1--C4 tools have zero exclusive CVEs, 
and most of their coverage consists entirely of common CVEs (e.g., CmpLog's 51 CVEs 
are all common), showing that traditional fuzzers largely overlap on the same vulnerabilities.

\begin{figure}[!htb]
    \centering
    \includegraphics[width=\columnwidth]{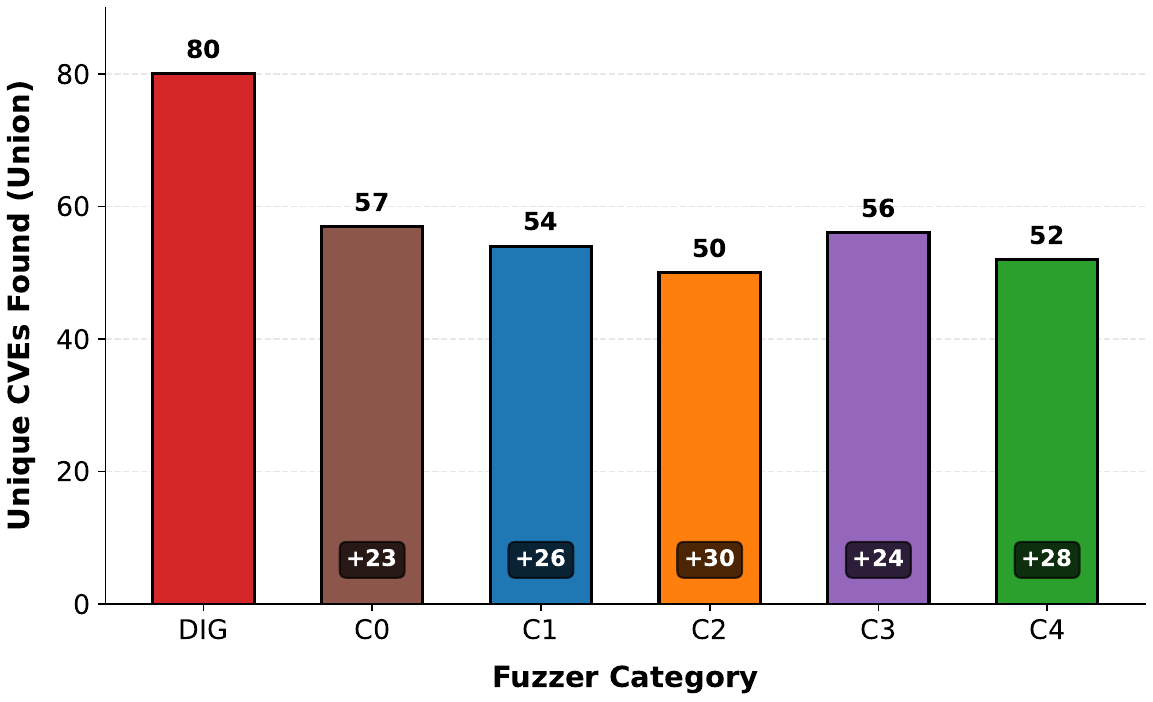}
    \caption{Unique CVEs found by the union of all tools in each category.}
    \label{fig:category_combined}
\end{figure}

To further understand how fuzzers from different categories perform in PoC 
generation, we compare DIG's CVE coverage against the union of all fuzzers in each category.
Figure~\ref{fig:category_combined} reveals that DIG alone outperforms the combined union 
of every category. Agentic systems (C0) together cover only 57 CVEs (23 fewer than 
DIG alone). Directed fuzzers (C2) collectively find only 50 CVEs (30 fewer), 
LLM-assisted input generation tools (C4) together find 52 CVEs (28 fewer), 
mutation-enhanced fuzzers (C3) find 56 CVEs (24 fewer), and 
baseline fuzzers (C1) find 54 CVEs (26 fewer). This shows that DIG's advantage is 
not attributable to the underperformance of any single baseline. Even when all tools 
within a category are combined to eliminate individual variance, DIG alone still 
achieves higher CVE coverage.

In addition, DIG's CVE coverage is also consistent across different file formats: DIG triggers all listed 
CVEs in PNG, SND, TIF, XML, Lua, PDF, and SQL, and triggers most listed CVEs in PHP 
(8 out of 10) and SSL (5 out of 7), showing that its effectiveness generalizes across 
diverse input formats and program sizes.

\subsubsection{Time-to-Event (TTE)}

\begin{figure}[!t]
    \centering
    \includegraphics[width=\columnwidth]{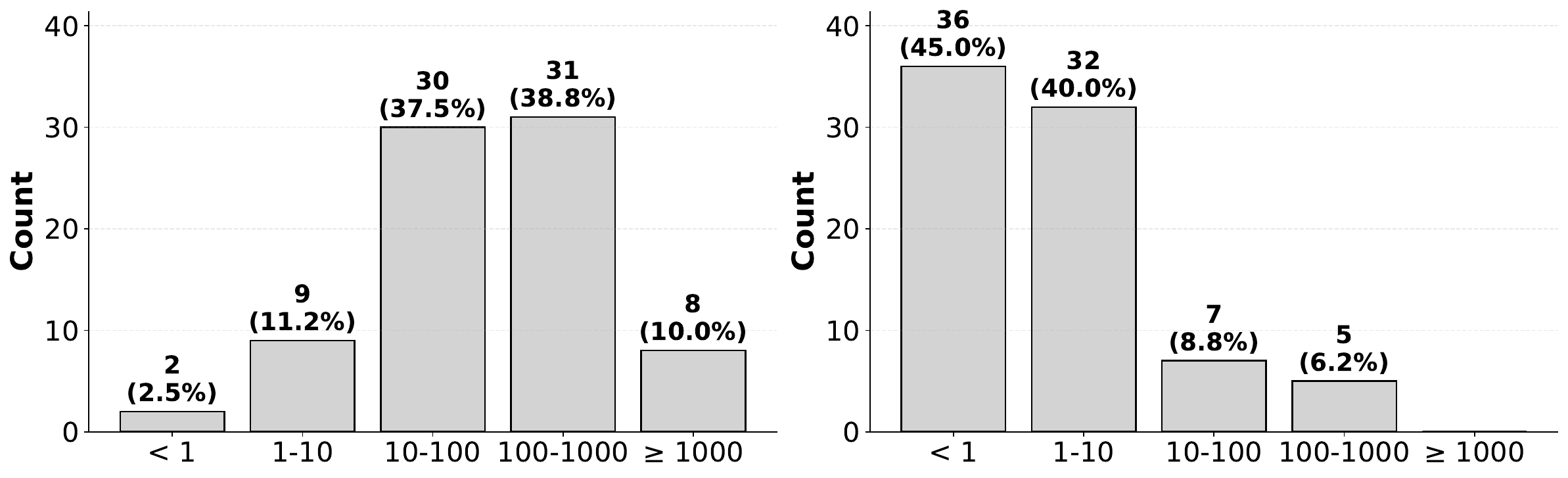}
    \caption{Distribution of DIG's speedup ranges. 
    Left: vs. average. Right: vs. best performer.}
    \label{fig:speedup_distribution}
\end{figure}

We also evaluate DIG's effectiveness by comparing its Time-to-Event (TTE) with that of
12 state-of-the-art tools. 
Table~\ref{tab:trigger} presents the detailed TTE results, where we compute 
two key metrics: (1) {vs. Avg.}, which measures DIG's speedup relative 
to the average TTE of all other tools, and (2) {vs. Best}, which 
measures DIG's speedup relative to the best-performing tool for each CVE.

DIG demonstrates strong TTE performance across the benchmark. 
Compared to the average performance of other tools, DIG is faster in 
92.9\% of cases (78 out of 84 CVEs), slower in 2 cases, and times out in the remaining 4 cases. 
Moreover, 48.8\% of cases show speedups exceeding 100×, and 10.0\% achieve 
speedups over 1000×. Even under the stricter comparison against the 
best-performing tool for each CVE, DIG remains faster in 52.4\% of cases 
(44 out of 84 CVEs), including 5 cases with speedups over 100×. 
Figure~\ref{fig:speedup_distribution} visualizes the distribution of these 
speedup ranges. The cases where DIG is slower than the best baseline are mostly 
easy-to-trigger targets. Among these 36 CVEs, the fastest baseline triggers 31 CVEs 
within 3 minutes. In 20 of these cases, DIG takes no more than 3 additional minutes 
compared with the fastest baseline. 
For example, on PNG003, the fastest baseline triggers the CVE in 
0.3 minutes, while DIG takes 1.5 minutes; this appears as a 5× slowdown, but the 
absolute gap is only 1.2 minutes. On these easy targets, seed-based, high-throughput 
fuzzers can quickly expose vulnerabilities close to existing seeds, and agentic systems 
such as PBFuzz and Cursor can also rapidly trigger them through direct code-level reasoning. 
DIG, by contrast, generates PoCs from scratch through 
generator evolution, which introduces additional overhead that is most apparent on 
easy-to-trigger CVEs.

In addition, DIG achieves remarkable speedups on several CVEs where even the 
best baseline struggles. On SQL001, DIG triggers the vulnerability in 1.4 minutes, 
outperforming the best baseline by 979.29× and the average by 1020.58×. 
Similarly, on SQL006, DIG achieves a 960.00× speedup over both the average and the 
best baseline. On PHP005, DIG triggers the CVE in 2.1 minutes while all other tools 
time out, yielding a 685.71× speedup.

\vspace{8pt}
\begin{tcolorbox}[colback=gray!10, colframe=black, boxrule=0.5pt, arc=2pt, left=6pt, right=6pt, top=6pt, bottom=6pt]
\textbf{RQ2 Takeaway:} DIG triggers 80 out of 138 CVEs (58.0\%), outperforming all 12 baselines 
and every category union, with 9 exclusive CVEs that no other tool can trigger. 
Compared to the average baseline TTE, DIG is faster in 78 out of 84 cases (92.9\%), 
with 8 cases exceeding 1000× speedup. While agentic systems (C0) demonstrate the potential
to trigger vulnerabilities that traditional fuzzers cannot, goal drift can limit their ability to reason about
hard-to-trigger vulnerabilities. DIG mitigates this limitation through oracle-guided generator evolution, 
which evaluates each iteration based on measurable constraint satisfaction and preserves 
generators that make progress toward the oracle, improving its ability to reason about hard-to-trigger vulnerabilities.
\end{tcolorbox}

\subsection{In-depth Analysis}

\subsubsection{Constraint Inference}

\begin{figure}[!htb]
    \centering
    \includegraphics[width=\columnwidth]{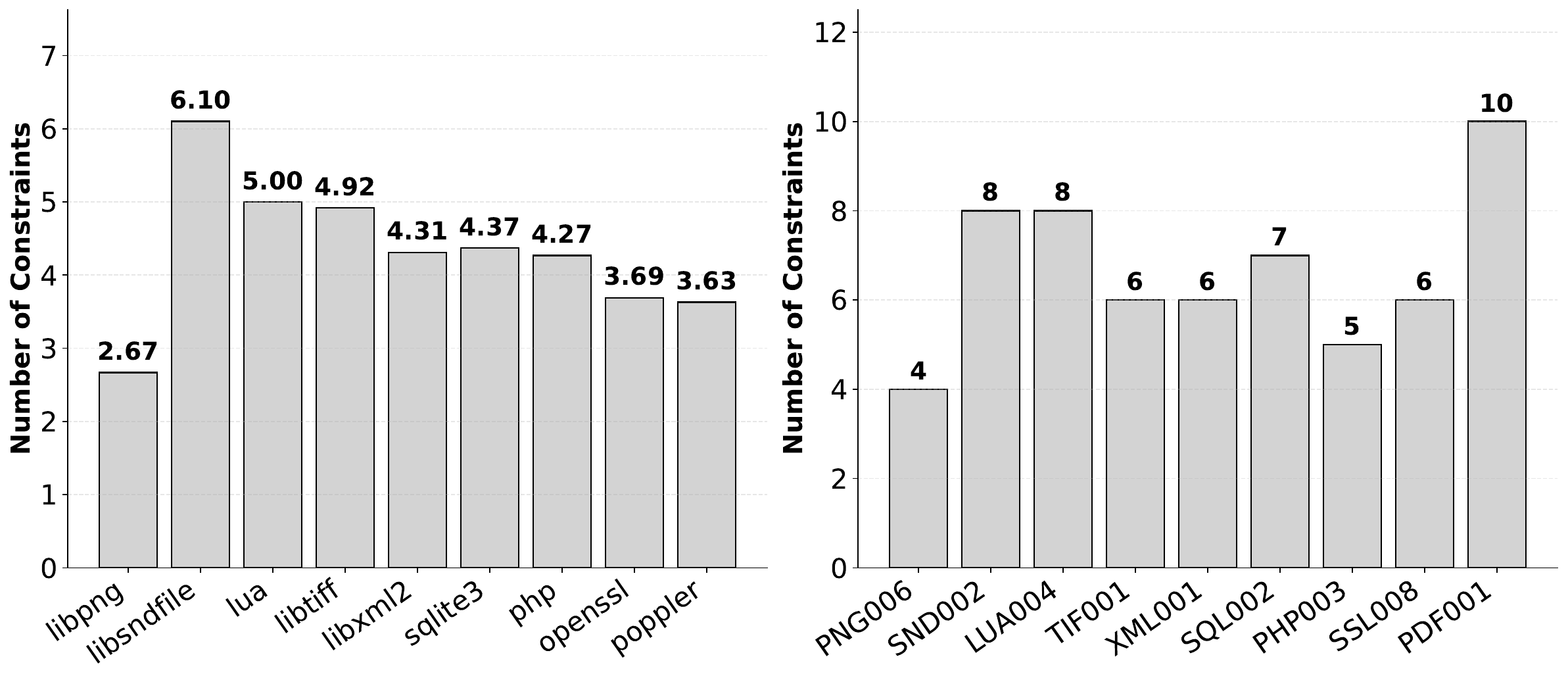}
    \caption{Constraint inference statistics. Left: average number of 
    inferred constraints across bugs in each program. Right: the bug with the most inferred constraints in each program.}
    \label{fig:constraint_inference}
\end{figure}

Figure~\ref{fig:constraint_inference} (left) shows the average number of oracle-related 
constraints inferred per vulnerability for each program. Across all programs, DIG infers an 
average of 4.28 constraints per vulnerability. The number of constraints varies across programs: 
libsndfile has the highest average (6.10), followed by Lua (5.00) and libtiff (4.92), 
while libpng has the lowest (2.67). Poppler (3.63) and OpenSSL (3.69) also have 
relatively few constraints per vulnerability.

Figure~\ref{fig:constraint_inference} (right) shows the vulnerability with the most inferred 
constraints in each program. Although the number of constraints does not directly determine constraint solving 
difficulty, it serves as a useful proxy for oracle complexity and correlates with 
DIG's TTE in practice. 
Among these hardest vulnerabilities, SND002 (8 constraints), PHP003 (5 constraints), and SSL008 
(6 constraints) are vulnerabilities that DIG fails to trigger within the time budget. For vulnerabilities that DIG 
successfully triggers, those with the most constraints tend to have the longest TTE 
within their program: LUA004 (8 constraints, 15.3 min) is the slowest among all Lua 
vulnerabilities; SQL002 (7 constraints, 60.1 min) is the slowest among all sqlite3 vulnerabilities; and 
PNG006 (4 constraints, 8.6 min) is the slowest among all libpng vulnerabilities. 
PDF001, also used as the motivating example in Section~\ref{sec:motivating_example_pdf001},
has the most constraints overall (10) and takes 110.7 minutes, making PDF001 the CVE that took the longest to trigger using 
generator evolution alone. These results suggest a 
positive correlation between the number of inferred constraints and DIG's TTE, as 
more constraints require more rounds of generator evolution to satisfy.

\subsubsection{Token Breakdown}

Regarding PoC generation token usage, Table~\ref{tab:cost_breakdown} reports 
the average LLM token usage per vulnerability for each program, broken down into
Constraint Inference (CI) and Generator Evolution (GE). Across all programs,
DIG consumes a total of 55.1M tokens, with CI accounting for 90.0\% (49.6M tokens) 
and GE for 10.0\% (5.5M tokens). 

The token disparity between CI and GE arises because CI requires extensive
agentic tool use to inspect source code, analyze calling relationships, and examine generator execution
traces before producing actionable evolution suggestions. 
This process often incurs many LLM API calls. In contrast, GE mainly
requires a small number of calls to update the generator.

\vspace{8pt}
\begin{table}[!htb]
    \centering
    \caption{Average LLM token usage (in thousands) per vulnerability for each program. 
    CI and GE denote Constraint Inference and Generator Evolution, respectively.}
    \label{tab:cost_breakdown}
    \footnotesize
    \begin{tabular}{lrrr}
    \toprule
    \textbf{Program} & \textbf{CI (K tok)} & \textbf{GE (K tok)} & \textbf{Total (K tok)} \\
    \midrule
    libpng     & 348.8   & 77.8    & 426.5 \\
    libsndfile & 1,709.3 & 241.5   & 1,950.8 \\
    libtiff    & 10,927.5 & 1,241.8 & 12,169.3 \\
    libxml2    & 6,861.4 & 452.8   & 7,314.1 \\
    lua        & 235.6   & 54.5    & 290.2 \\
    openssl    & 10,931.1 & 1,234.7 & 12,165.8 \\
    php        & 1,848.7 & 265.6   & 2,114.3 \\
    poppler    & 12,670.6 & 1,525.4 & 14,195.9 \\
    sqlite3    & 4,040.5 & 410.4   & 4,450.9 \\
    \midrule
    \textbf{TOTAL} & \textbf{49,573.5} & \textbf{5,504.4} & \textbf{55,077.9} \\
    \bottomrule
    \end{tabular}
\end{table}

\subsubsection{Cost and LLM Model Comparison}
\begin{figure*}[t]
    \centering
    \captionsetup[subfigure]{labelfont=normalfont, labelformat=simple, skip=2pt}
    \renewcommand{\thesubfigure}{(\alph{subfigure})}
    \begin{subfigure}{0.32\textwidth}
        \centering
        \includegraphics[width=\linewidth]{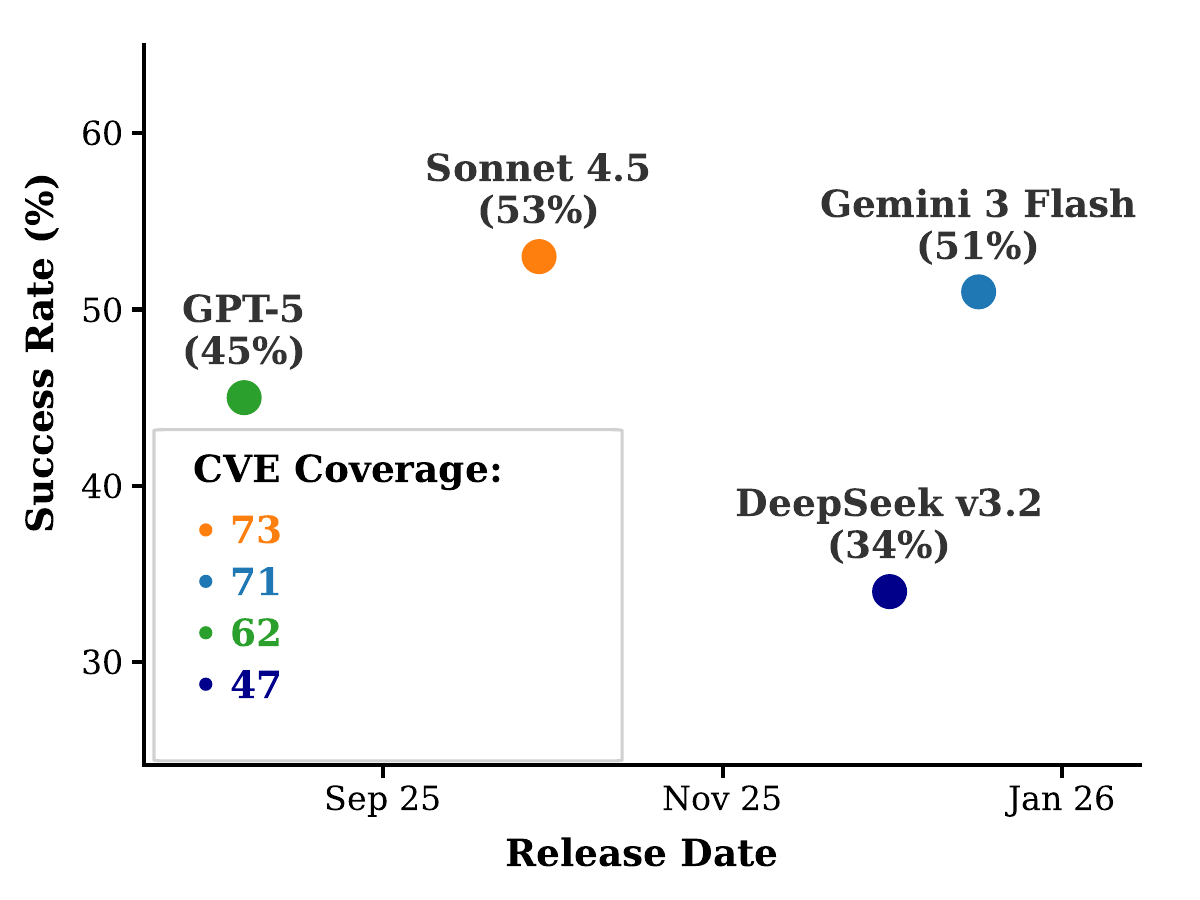}
        \caption{}
        \label{fig:llm_date_sr}
    \end{subfigure}
    \hfill
    \begin{subfigure}{0.32\textwidth}
        \centering
        \includegraphics[width=\linewidth]{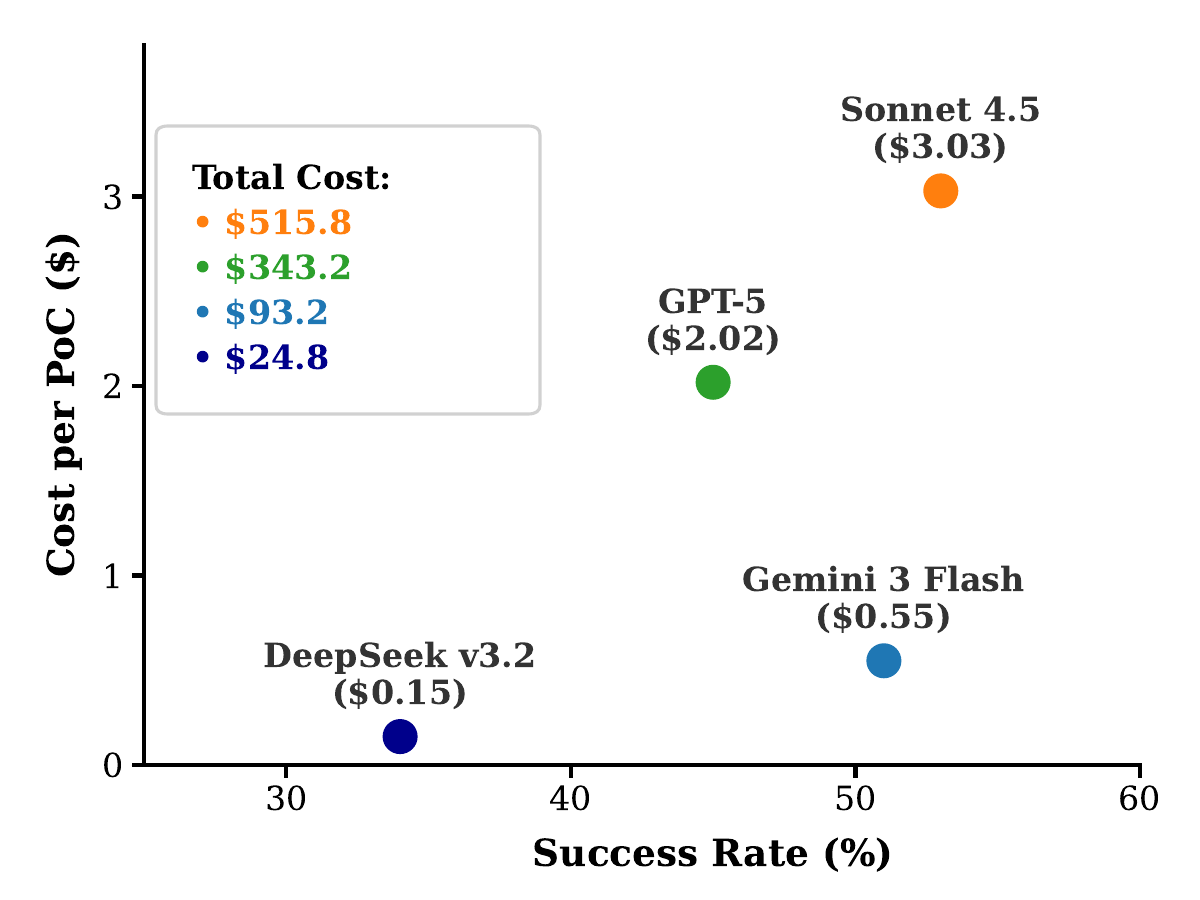}
        \caption{}
        \label{fig:llm_cost_sr}
    \end{subfigure}
    \hfill
    \begin{subfigure}{0.32\textwidth}
        \centering
        \includegraphics[width=\linewidth]{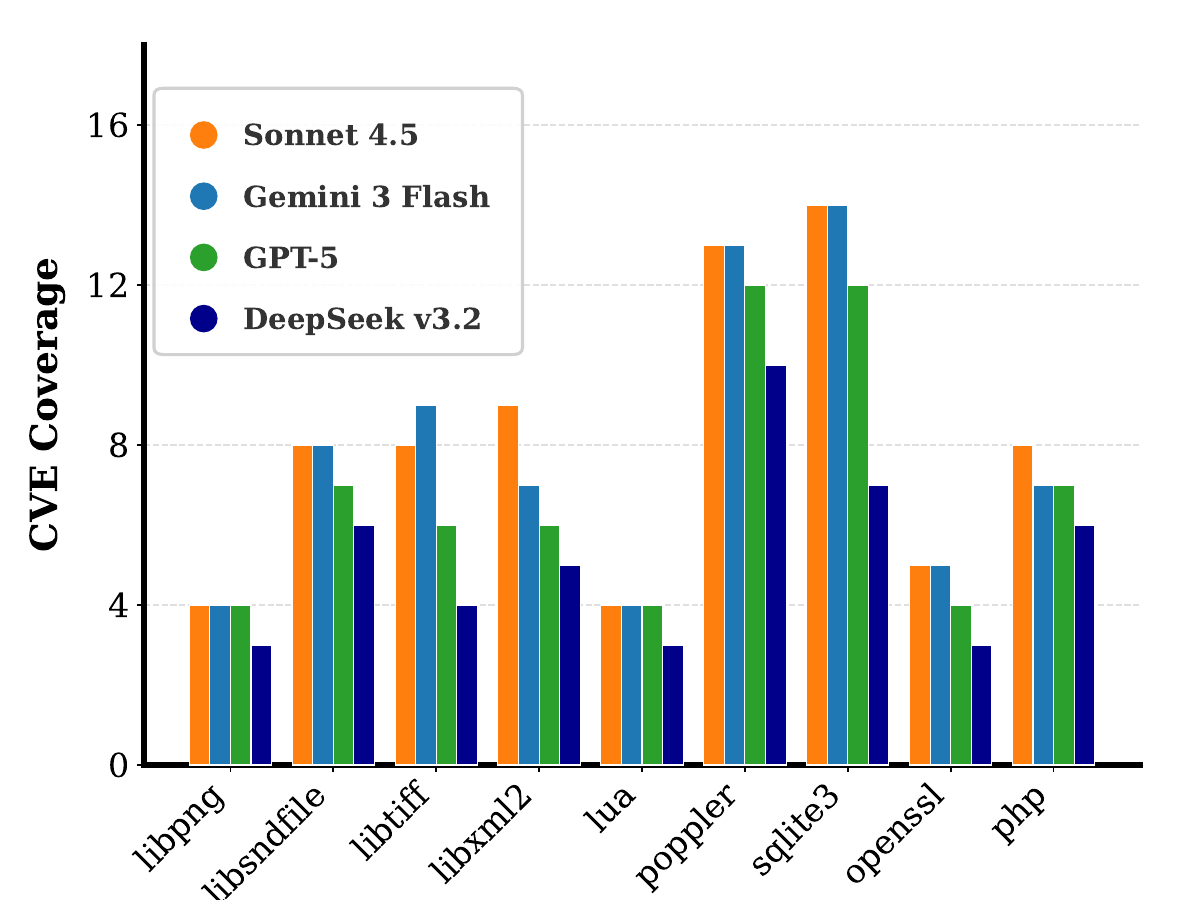}
        \caption{}
        \label{fig:llm_cve_projects}
    \end{subfigure}
    \vspace{8pt}
    \caption{Comparison of DIG$^\dag$ with different LLM backends: 
    (a) success rate and model release date, 
    (b) cost per PoC generation and success rate, and 
    (c) CVE coverage across different open-source projects.}
    \label{fig:llm_model_comparison}
\end{figure*}

The total cost of oracle synthesis is \$16.2 for 135 CVEs, corresponding to an average of \$0.12 per oracle.
We also evaluate the success rate of PoC generation (generator evolution alone) and its cost for four state-of-the-art LLMs: 
GPT-5~\cite{singh2025openai}, Sonnet 4.5~\cite{anthropic2025sonnet45},
Gemini 3 Flash~\cite{google2025gemini3flash}, and DeepSeek v3.2~\cite{liu2025deepseek},
where DeepSeek v3.2 is an open-weight model and the remaining models are closed-source models.
Figure~\ref{fig:llm_model_comparison}\subref{fig:llm_date_sr} plots each model's success rate and release date.
Although the earliest and latest releases differ by
nearly four months, we do not observe a strong correlation between model release
date and PoC success rate.

Overall, the closed-source models achieve higher PoC success rates than
the open-weight models.
Among these models, Sonnet 4.5 achieves the highest success rate
(53\%), followed by Gemini 3 Flash (51\%) and GPT-5 (45\%), while the open-weight model
DeepSeek v3.2 attains a success rate of 34\%.
On average, the closed-source models achieve success rates that are 15.7
percentage points higher than those of the open-weight model, corresponding to a
46.1\% relative difference.

For PoC generation cost, we evaluate four LLM backends on 170 CVE instances.\footnote{This 
differs from the 135 CVEs in oracle synthesis because some projects contain 
multiple programs, and some CVEs are not compiled in Magma's binary configurations.}
Figure~\ref{fig:llm_model_comparison}\subref{fig:llm_cost_sr} shows substantial cost 
variation: DeepSeek v3.2 (\$0.15 per PoC, \$24.8 total), Gemini 3 Flash (\$0.55, \$93.2), 
GPT-5 (\$2.02, \$334.2), and Sonnet 4.5 (\$3.03, \$515.8). Success rates also 
vary: DeepSeek v3.2 (28\%), GPT-5 (50\%), Gemini 3 Flash (55\%), and Sonnet 4.5 (58\%). 
Gemini 3 Flash emerges as the most cost-effective option, achieving 55\% success 
at \$93.2---5.5× cheaper than Sonnet 4.5 (58\%, \$515.8) with only a 3-percentage-point lower 
success rate. Sonnet 4.5 offers the highest performance but at a premium cost. 
DeepSeek v3.2, while cheapest (\$24.8), delivers a poor success rate (28\%). GPT-5 (\$334.2, 50\%)
falls short on both dimensions: it underperforms Gemini 3 Flash 
in success rate (50\% vs. 55\%) while costing 3.6× more.

Figure~\ref{fig:llm_model_comparison}\subref{fig:llm_cve_projects} presents a per-project comparison of CVE
coverage achieved by each LLM on the Magma benchmark.
Sonnet 4.5 and Gemini 3 Flash achieve the highest CVE coverage across all
projects.
Among closed-source models, Sonnet 4.5 and Gemini 3 Flash cover all CVEs found by GPT-5,
and additionally cover 11 unique vulnerabilities that GPT-5 cannot find.
Among these three models, Sonnet 4.5 finds three unique vulnerabilities, while Gemini 3
Flash finds one unique vulnerability.

\vspace{6pt}
\begin{table}[!htb]
    \centering
    \caption{TTE (in minutes) across 5 independent runs per vulnerability. T.O.\ denotes timeout.}
    \label{tab:consistency}
    \footnotesize
    \setlength{\tabcolsep}{4pt}
    \begin{tabular}{lrrrrrrc}
    \toprule
    \textbf{Bug ID} & \textbf{Run 1} & \textbf{Run 2} & \textbf{Run 3} & \textbf{Run 4} & \textbf{Run 5} & \textbf{Std} & \textbf{Succ.} \\
    \midrule
    LUA001 & 13.7 & 1.6 & 1.5 & 1.5 & 6.5 & 5.3 & 5/5 \\
    LUA002 & 1.7 & 3.7 & 1.6 & 1.6 & 1.6 & 0.9 & 5/5 \\
    PDF005 & 34.0 & 23.0 & 24.8 & T.O. & 29.2 & 4.9 & 4/5 \\
    PDF022 & 1.4 & 1.5 & 1.3 & 3.4 & 1.6 & 0.9 & 5/5 \\
    PHP002 & 1.2 & 1.2 & 1.3 & 1.1 & 1.2 & 0.1 & 5/5 \\
    PHP009 & 2.8 & 1.6 & 6.6 & 3.6 & 2.9 & 1.9 & 5/5 \\
    PNG001 & 2.2 & 2.1 & 0.9 & 1.8 & 2.0 & 0.5 & 5/5 \\
    PNG007 & 1.4 & 3.0 & 1.6 & 1.7 & 1.8 & 0.6 & 5/5 \\
    SND016 & 3.5 & 38.4 & 31.8 & 38.7 & 7.1 & 17.2 & 5/5 \\
    SND020 & 5.7 & 4.3 & 3.8 & 4.3 & 6.9 & 1.3 & 5/5 \\
    SQL003 & 0.6 & 11.3 & 0.8 & 0.8 & 1.5 & 4.7 & 5/5 \\
    SQL018 & 17.9 & 6.5 & 26.3 & 1.7 & 5.2 & 10.3 & 5/5 \\
    SSL001 & 45.6 & 46.3 & 45.4 & 53.0 & 2.1 & 20.6 & 5/5 \\
    SSL003 & 0.9 & 0.9 & 0.9 & 1.0 & 1.0 & 0.1 & 5/5 \\
    TIF008 & 11.0 & 12.5 & 16.7 & 62.6 & 52.8 & 24.6 & 5/5 \\
    TIF014 & 26.6 & 4.1 & 4.3 & 1.0 & 3.5 & 10.5 & 5/5 \\
    XML001 & 1.8 & 43.7 & 3.5 & 3.9 & 10.2 & 17.7 & 5/5 \\
    XML010 & 28.5 & 9.5 & T.O. & T.O. & T.O. & 13.4 & 2/5 \\
    \bottomrule
    \end{tabular}
\end{table}

\subsubsection{Consistency of PoC Generation}
In our main evaluation setting, we ran DIG once per target program due to cost constraints.
To understand the consistency of PoC generation, we randomly selected two vulnerabilities from
each program and ran DIG five times.
Table~\ref{tab:consistency} reports the TTE for each run along with the standard deviation and success count.
DIG successfully triggers vulnerabilities in 86 out of 90 runs (95.6\%), with 
16 out of 18 bugs achieving a 5/5 success rate. 
Among the 18 bugs, 8 exhibit low variance (std $<$ 2 min), 
indicating highly stable generation for vulnerabilities with simpler constraint structures 
(e.g., PHP002 with std = 0.1 min, SSL003 with std = 0.1 min).
The remaining bugs show higher variance due to the stochastic nature of LLM-guided 
generator evolution, where different runs may explore different constraint-solving 
strategies. Despite this variance, DIG consistently triggers these vulnerabilities 
across runs, confirming that the overall approach is reliable even when individual 
run times fluctuate.

\subsection{Ablation Study}
To understand how oracle-guided generator evolution and oracle-guided directed
mutation contribute to DIG’s performance, we conduct an ablation study with two variants.
DIG$^\dag$ removes low-level directed mutation and retains only generator evolution, 
isolating the contribution of high-level generation. DIG$^\ddag$ removes generator evolution 
by disabling the agentic tool set, including static analysis, dynamic analysis,
and source-code inspection.
As a result, DIG$^\ddag$ becomes a prompt-based input generation variant, 
similar to G$^2$Fuzz, while still keeping oracle-guided directed mutation enabled.
We evaluate both variants against full DIG. We randomly select two vulnerabilities from each program
that DIG can trigger. Table~\ref{tab:ablation} presents the results.

\par\vspace{6pt}
\begin{table}[!htbp]
    \centering
    \caption{Ablation study results showing the contribution of oracle-guided generator evolution and directed mutation. \checkmark indicates successful PoC generation; $\times$ indicates failure.}
    \label{tab:ablation}
    {\scriptsize
    \setlength{\tabcolsep}{3pt}
    \begin{tabular}{l|cccccccccccccccccc}
    \toprule
    \textbf{Variant} & \rotatebox{90}{\textbf{PNG001}} & \rotatebox{90}{\textbf{PNG006}} & \rotatebox{90}{\textbf{XML006}} & \rotatebox{90}{\textbf{XML012}} & \rotatebox{90}{\textbf{PDF004}} & \rotatebox{90}{\textbf{PDF014}} & \rotatebox{90}{\textbf{SND006}} & \rotatebox{90}{\textbf{SND020}} & \rotatebox{90}{\textbf{TIF008}} & \rotatebox{90}{\textbf{TIF014}} & \rotatebox{90}{\textbf{SQL002}} & \rotatebox{90}{\textbf{SQL014}} & \rotatebox{90}{\textbf{LUA001}} & \rotatebox{90}{\textbf{LUA004}} & \rotatebox{90}{\textbf{SSL009}} & \rotatebox{90}{\textbf{SSL016}} & \rotatebox{90}{\textbf{PHP001}} & \rotatebox{90}{\textbf{PHP005}} \\
    \midrule
    \textbf{DIG} & \checkmark & \checkmark & \checkmark & \checkmark & \checkmark & \checkmark & \checkmark & \checkmark & \checkmark & \checkmark & \checkmark & \checkmark & \checkmark & \checkmark & \checkmark & \checkmark & \checkmark & \checkmark \\
    \textbf{DIG$^\dag$} & \checkmark & \checkmark & \checkmark & $\times$ & $\times$ & $\times$ & \checkmark & \checkmark & $\times$ & \checkmark & $\times$ & \checkmark & \checkmark & \checkmark & \checkmark & \checkmark & \checkmark & \checkmark \\
    \textbf{DIG$^\ddag$} & \checkmark & $\times$ & $\times$ & \checkmark & $\times$ & $\times$ & $\times$ & \checkmark & \checkmark & \checkmark & \checkmark & $\times$ & $\times$ & \checkmark & $\times$ & $\times$ & $\times$ & $\times$ \\
    \bottomrule
    \end{tabular}
    }
\end{table}

The results reveal the distinct contributions of each component.
\textbf{DIG} successfully triggers all 18 CVEs (100\%).
\textbf{DIG$^\dag$} (no directed mutation) triggers 13/18 (72.2\%), 
failing on 5 CVEs (XML012, PDF004, PDF014, TIF008, SQL002). 
These failures suggest that generator evolution alone may leave
the oracle unsatisfied when triggering depends on hidden state relationships
or hard-to-infer runtime constraints. Oracle-guided directed mutation complements
this by performing high-throughput local search over oracle-relevant input regions,
using branch-distance feedback to move candidate inputs closer to satisfying the oracle
without explicitly inferring all missing constraints.

\textbf{DIG$^\ddag$} (no generator evolution) drops further to 7/18 (38.9\%), 
failing on 11 CVEs. This 61.1\% drop in success rate demonstrates that 
generator evolution is the primary driver of DIG's effectiveness:
without systematic constraint reasoning, the LLM may generate PoCs blindly.
Notably, SQL002 and TIF008 succeed in DIG$^\ddag$ but fail in DIG$^\dag$, 
indicating that for these CVEs the vulnerability can be triggered through 
directed mutations with prompt-based input generation.

\vspace{8pt}
\begin{tcolorbox}[colback=gray!10, colframe=black, boxrule=0.5pt, arc=2pt, left=6pt, right=6pt, top=6pt, bottom=6pt]
\textbf{RQ4 Takeaway:} 
Oracle-guided generator evolution is the primary driver of DIG's effectiveness,  
while oracle-guided directed mutation provides a complementary boost.
\end{tcolorbox} 

\subsection{New Vulnerability Finding}

While DIG targets one-day vulnerabilities, its techniques can extend to zero-day 
vulnerability discovery. We demonstrate this potential by using LLMs to scan code for
potentially vulnerable code locations and DIG to verify whether vulnerabilities actually exist. 
We evaluate DIG on the same Magma benchmark programs used in our one-day evaluation,
limiting the study to libpng and libsndfile due to budget constraints.
The libpng and libsndfile programs are widely deployed libraries 
that have undergone extensive testing, making them challenging 
targets for zero-day vulnerability finding.
However, the results are encouraging. We uncovered 6 previously unknown vulnerabilities.

Table~\ref{tab:new_bugs} summarizes the discovered vulnerabilities. The libpng 
vulnerability (Bug \#1) is a high-severity stack buffer overflow in \texttt{png\_do\_bgr()} 
caused by integer overflow when processing malicious PNG files with extreme width 
values. The 5 libsndfile bugs span 
multiple severity levels: 3 critical bugs (heap overflow, stack 
overflow, and NULL pointer dereference) and 2 high-severity integer overflows. 

\par\vspace{6pt}
\begin{table}[!htbp]
    \centering
    \caption{New vulnerabilities discovered by DIG in the latest versions of libpng and libsndfile.}
    \label{tab:new_bugs}
    {\scriptsize
    \setlength{\tabcolsep}{3pt}
    \renewcommand{\arraystretch}{1.05}
    \begin{tabularx}{\columnwidth}{l l l X l}
    \toprule
    \textbf{ID} & \textbf{Library} & \textbf{Type} & \textbf{Location} & \textbf{Severity} \\
    \midrule
    \#1 & libpng & Stack Overflow & \texttt{png\_do\_bgr()} & High \\
    \midrule
    \#2 & libsndfile & Heap Overflow & \texttt{psf\_save\_write\_chunk()} & Critical \\
    \#3 & libsndfile & Stack Overflow & \texttt{let2i\_array()} & Critical \\
    \#4 & libsndfile & NULL Deref & \texttt{sf\_open\_virtual()} & Critical \\
    \#5 & libsndfile & Int Overflow & \texttt{wav\_write\_fmt\_chunk()} & High \\
    \#6 & libsndfile & Int Overflow & \texttt{wavex\_write\_fmt\_chunk()} & High \\
    \bottomrule
    \end{tabularx}
    }
\end{table}

To understand how DIG discovers vulnerabilities, we present a case study of Bug \#1, 
a stack buffer overflow in libpng's \texttt{png\_do\_bgr()} function. 
When a malicious PNG file specifies an extreme width (e.g., \texttt{0xFFFFFFFF}), 
the calculation \texttt{rowbytes = width * 3} overflows, wrapping to 
\texttt{0xFFFFFFFD} on 32-bit systems. This causes the program to allocate a small stack 
buffer while accessing it based on the original large width, triggering overflow.
DIG first synthesized an oracle checking for integer overflow in the rowbytes 
calculation: \texttt{(width > UINT32\_MAX / channels)}. DIG then inferred the 
constraints related to the oracle and found the following conditions: (1) the PNG width is set to an extreme value 
($\geq$ \texttt{0x55555556}), (2) the color type is \texttt{PNG\_COLOR\_TYPE\_RGB} 
(3 channels), (3) the bit depth is 8, and (4) the application uses \texttt{PNG\_TRANSFORM\_BGR}. 
These constraints guided DIG to generate inputs toward extreme width values 
combined with BGR transformation.

\section{Discussion}
\label{sec: discussion}

\subsection{Generality}
DIG’s approach is broadly applicable. First, DIG is agnostic to vulnerability classes,
such as buffer overflows, use-after-free bugs, and integer overflows. 
Our large-scale evaluation on Magma shows that DIG can trigger diverse
types of vulnerabilities. Second, DIG does not require formal grammars
or manually specified input formats, making it applicable to both text-based formats,
such as XML and SQL, and binary formats, such as PDF and TIFF. 
Third, although DIG currently targets C/C++, its LLM-based reasoning can
naturally extend to other languages. DIG’s success on Poppler, which uses
complex C++ features such as polymorphism and dynamic dispatch, suggests
potential applicability to similar object-oriented languages.

\subsection{Limitations}

\subsubsection{Assumptions on Vulnerability Artifacts}
DIG targets one-day vulnerabilities, assuming availability of 
vulnerability-related artifacts such as patches, vulnerability reports, or 
candidate vulnerable functions. These artifacts provide initial guidance for 
oracle synthesis and drive input generation. This design choice aligns with 
the one-day vulnerability setting where patches and vulnerability details are publicly 
released, distinguishing it from zero-day discovery where such information 
is unavailable.

\subsubsection{Training Data Contamination}
Magma CVEs may appear in the training data of state-of-the-art LLMs, which is a common
challenge for LLM-based vulnerability research on public benchmarks. However, DIG’s advantages
in CVE coverage and TTE over other LLM-based approaches under the same model setting suggest
that its gains come from its design and cannot be explained by training data exposure alone.

\section{Related Work}

\paragraph{Directed Greybox Fuzzing}
AFLGo~\cite{bohme2017directed} pioneered 
control-flow distance to guide execution toward target locations.  
Hawkeye~\cite{chen2018hawkeye} improves the control-flow distance metric by address-
ing challenges related to indirect calls and incorporating
call trace similarity for more accurate targeting.
WindRanger~\cite{du2022windranger} considers deviated basic blocks that steer 
execution away from targets. Lyso~\cite{bao2025alarms} enhances distance metrics 
with multi-step guidance. Beacon~\cite{huang2022beacon} uses static 
analysis to identify unreachable code regions and inserts assertion checks. 
SieveFuzz~\cite{srivastava2022one} dynamically prunes unreachable paths at runtime. 
SelectFuzz~\cite{luo2023selectfuzz} and DAFL~\cite{kim2023dafl} apply selective instrumentation to exclude 
control- and data-independent code regions from coverage tracking.

\paragraph{LLM-Assisted Testcase Generation}
Several approaches use LLMs to generate initial 
seeds or mutate existing ones. Fuzz4All~\cite{xia2024fuzz4all} synthesizes and 
refines input-generation prompts based on coverage feedback. G²fuzz~\cite{zhang2025low} 
and ELFuzz~\cite{chen2025elfuzz} synthesize generators and evolve them through 
mutation. ChatFuzz and LlamaFuzz~\cite{zhang2024llamafuzz} provide structure-aware 
mutation operators. For protocol fuzzing, ChatAFL~\cite{meng2024large} extracts 
message structures from specifications to guide stateful sequence generation, while 
mGPTFuzz~\cite{ma2024mgptfuzz} mines protocol standards to derive state models.

Complementary work applies LLMs to enhance 
symbolic execution. AutoBug~\cite{li2025large} uses LLMs to replace SMT solvers, 
representing path constraints as code for direct reasoning. COTTONTAIL~\cite{tu2025large} 
leverages LLMs for structure-aware constraint selection and concolic execution. 
CONCOLLMIC~\cite{luo2026agentic} employs LLM agents for language-agnostic symbolization. 
These approaches target code coverage rather than vulnerability reproduction.

\section{Conclusion}
We present DIG, an oracle-guided PoC generation system for one-day vulnerabilities.
DIG introduces oracle synthesis, oracle-guided generator evolution, and oracle-guided directed mutation to
identify necessary constraints for vulnerability triggering and solve them effectively.
Extensive evaluation on the Magma benchmark shows
that DIG substantially outperforms state-of-the-art agents and fuzzers,
achieving higher CVE coverage, faster time to trigger the vulnerabilities,
and discovering 9 exclusive vulnerabilities not triggered by existing techniques. We also demonstrate
DIG's potential for zero-day vulnerability discovery and find 6 new vulnerabilities.

\section*{Ethical Considerations}
The primary ethical consideration in our work is ensuring that the
vulnerabilities we identify and the associated proof-of-concept exploits are
responsibly disclosed.
We have followed standard practices for responsible disclosure by notifying the
affected parties of any vulnerabilities discovered during our experiments before
any public release of the findings.
This ensures that the software vendors have the opportunity to address the
vulnerabilities before they are exposed to the public.
Additionally, our work does not involve any experiments on human subjects, use
of personal data, or other activities that could raise significant ethical
concerns.

\FloatBarrier

\bibliographystyle{unsrt}
\bibliography{bib}

\appendix
\subsection{Evaluation Baselines and Benchmark}
\label{sec:appendix_baselines}

\begin{table}[H]
    \centering
    \caption{Overview of tools used for evaluation.}
    \label{tab:fuzzing_baselines}
    {\scriptsize
    \setlength{\tabcolsep}{3pt}
    \renewcommand{\arraystretch}{1.05}
    \newcolumntype{F}{>{\raggedright\arraybackslash}p{2.2cm}}
    \newcolumntype{C}{>{\centering\arraybackslash}p{0.6cm}}
    \begin{tabularx}{\columnwidth}{C F X}
    \toprule
    \textbf{Cat.} & \textbf{Fuzzer} & \textbf{Description} \\
    \midrule
    \multirow{2}{*}{C0} &
    PBFuzz~\cite{zeng2025pbfuzz} \newline
    Cursor~\cite{anysphere2023cursor} &
    Agentic PoC generation \newline
    General-purpose coding agent \\
    \midrule
    \multirow{2}{*}{C1} &
    AFL~\cite{zalewski_afl} \newline
    AFL++~\cite{fioraldi2020afl++} &
    Baseline \newline
    Community-enhanced baseline \\
    \midrule
    \multirow{3}{*}{C2} &
    AFLGo~\cite{bohme2017directed}\newline
    SelectFuzz~\cite{luo2023selectfuzz} \newline
    Titan~\cite{huang2024titan} &
    Distance-based seed scheduling \newline
    Selective path exploration \newline
    Multi-target synergy \\
    \midrule
    \multirow{2}{*}{C3} &
    CmpLog~\cite{aschermann2019redqueen} \newline
    FOX~\cite{she2024fox} &
    Comparison logging \newline
    Online stochastic control \\
    \midrule
    \multirow{3}{*}{C4} &
    G$^2$Fuzz~\cite{zhang2025low} \newline
    LLAMAFUZZ~\cite{zhang2024llamafuzz} \newline
    SeedAIchemy~\cite{wen2025seedaichemy} &
    LLM-synthesized input generator \newline
    LLM-assisted mutation \newline
    LLM-assisted seed corpus generation \\
    \bottomrule
    \end{tabularx}
    }
\end{table}

\begin{table}[H]
    \centering
    \caption{Magma targets with approximate size (LoC), CVE counts, input formats and types.}
    \label{tab:magma_target_size}
    {\scriptsize
    \setlength{\tabcolsep}{3pt}
    \renewcommand{\arraystretch}{1.05}
    \begin{tabular}{c c c >{\centering\arraybackslash}p{3.6cm} c}
    \toprule
    \textbf{Target} & \textbf{LoC} & \textbf{CVE Count} & \textbf{Format} & \textbf{Type} \\
    \midrule
    libpng     & 95k  & 7  & png                 & Binary \\
    libsndfile & 83k  & 18 & wav, aiff, au, caf  & Binary \\
    libtiff    & 95k  & 14 & tiff                & Binary \\
    libxml2    & 320k & 17 & xml                 & Text \\
    Lua        & 21k  & 4  & lua                 & Text \\
    poppler    & 260k & 22 & pdf                 & Binary \\
    SQLite     & 387k & 20 & sql                 & Text \\
    OpenSSL    & 630k & 20 & cryptography        & Binary \\
    PHP        & 1.6M & 16 & php, json, exif, jpeg    & Text and Binary \\
    \bottomrule
    \end{tabular}
    }
\end{table}

\begin{table*}[!ht]
    \centering
    \caption{
    TTE, measured in minutes, for each CVE in the Magma benchmark.
    T.O. indicates that the fuzzer failed to generate a PoC within the time budget.
    $\varnothing$ denotes missing data, and N.A. denotes that fuzzer deployment is not applicable.
    }
    \label{tab:trigger}
    {\scriptsize
    \setlength{\tabcolsep}{1.2pt}
    \renewcommand{\arraystretch}{0.87}
    \newcolumntype{C}{>{\centering\arraybackslash}X}
    \begin{tabularx}{\textwidth}{l|*{13}{C}|C|C}
    \toprule
    \textbf{Vul ID} & \textbf{DIG} & \textbf{PBFuzz} & \textbf{Cursor} & \textbf{\shortstack[c]{G$^2$fuzz}} & \textbf{\shortstack[c]{LLAMA\\FUZZ\textsuperscript{*}}} & \textbf{\shortstack[c]{SeedAI\\chemy\textsuperscript{*}}} & \textbf{\shortstack[c]{AFL\\Go}} & \textbf{\shortstack[c]{Select\\fuzz}} & \textbf{Titan} & \textbf{AFL} & \textbf{AFL++} & \textbf{\shortstack[c]{CmpLog}} & \textbf{FOX} & \textbf{\shortstack[c]{vs.\\Avg.}} & \textbf{\shortstack[c]{vs.\\Best}} \\
    \midrule
PNG001 & 2.8 & 10.8 & 2.0 & 12.6 & T.O. & T.O. & 1422.2 & T.O. & T.O. & 1208.7 & 1261.7 & 1416.3 & T.O. & $373.04\times$ & $0.71\times$\\
PNG003 & 1.5 & 6.7 & 5.6 & 5.7 & 0.4 & 2.0 & 0.3 & 0.3 & 0.3 & 0.3 & 0.3 & 0.3 & 0.3 & $1.25\times$ & $0.20\times$\\
PNG006 & 8.6 & 4.5 & T.O. & T.O. & 0.3 & 3.0 & T.O. & T.O. & T.O. & T.O. & T.O. & 0.3 & 46.3 & $98.20\times$ & $0.03\times$\\
PNG007 & 1.4 & 11.9 & 7.5 & 121.9 & 300.0 & 300.0 & 55.4 & 87 & 209 & 407.7 & 46.4 & 41.8 & 77.1 & $99.15\times$ & $5.36\times$\\
SND001 & 1.6 & 1.4 & 2.0 & T.O. & 1.0 & 540.0 & 2.9 & 2.2 & 10 & 4.3 & 6.4 & 12.6 & 1.9 & $105.45\times$ & $0.62\times$\\
SND005 & 15.1 & 4.0 & 7.8 & T.O. & 16.0 & 420.0 & 0.5 & 0.6 & 1.2 & 0.8 & 41.8 & 63.5 & 2.5 & $11.03\times$ & $0.03\times$\\
SND006 & 10.3 & T.O. & T.O. & T.O. & 5.0 & 9.0 & 3.9 & 2.9 & 441 & 21.7 & 61.5 & 168.9 & 71.8 & $41.31\times$ & $0.28\times$\\
SND007 & 8.4 & T.O. & T.O. & T.O. & 8.0 & 11.0 & 3.7 & 2.9 & 11.7 & 5.4 & 57.9 & 137.1 & 77.5 & $45.98\times$ & $0.35\times$\\
SND016 & 22.5 & 6.4 & T.O. & T.O. & T.O. & 41.0 & T.O. & T.O. & T.O & T.O. & T.O. & T.O. & 326.5 & $49.38\times$ & $0.28\times$\\
SND017 & 1.6 & 3.3 & 4.7 & 26.2 & 7.0 & 13.0 & 17.7 & 12.3 & 9.8 & 19.5 & 72.8 & 69.3 & 0.7 & $13.35\times$ & $0.44\times$\\
SND020 & 0.8 & 2.0 & 4.5 & 88.7 & 23.0 & 20.0 & 32.3 & 222.9 & 173.9 & 69.8 & 120.8 & 159.8 & 18.1 & $97.48\times$ & $2.50\times$\\
SND024 & 3.8 & T.O. & T.O. & T.O. & 1.0 & 8.0 & 3.5 & 2.8 & 11.7 & 5.3 & 13.3 & 43.3 & 26.5 & $97.27\times$ & $0.26\times$\\
TIF001 & 38.5 & T.O. & T.O. & T.O. & T.O. & T.O. & T.O. & T.O. & T.O. & T.O. & 1415.1 & T.O. & T.O. & $37.35\times$ & $36.76\times$\\
TIF002 & 43.3 & T.O. & T.O. & 168.3 & T.O. & 480.0 & 895.3 & 961.4 & 1312.3 & 622.9 & 394.1 & 513.1 & 426.7 & $19.43\times$ & $3.89\times$\\
TIF005 & 1.0 & 9.7 & 3.8 & T.O. & 15.0 & 5.0 & T.O. & T.O. & T.O. & T.O. & T.O. & 103.8 & 56.2 & $736.12\times$ & $3.80\times$\\
TIF006 & 3.0 & 1.9 & 11.0 & 48.3 & 12.0 & 9.0 & 997.5 & 1305.3 & 933.3 & 836 & 669.1 & 100.6 & 133.3 & $140.48\times$ & $0.63\times$\\
TIF007 & 3.1 & 6.8 & 2.7 & 3.8 & 0.5 & 1.0 & 0.5 & 0.5 & 5.8 & 0.5 & 0.3 & 0.3 & 2.2 & $0.67\times$ & $0.10\times$\\
TIF008 & 99.8 & 7.6 & T.O. & 537.5 & 2880.0 & 660.0 & 1140.2 & 1345.7 & 1314.4 & 1016.3 & 1241.2 & 1084.2 & 1374.4 & $11.72\times$ & $0.08\times$\\
TIF009 & 13.3 & 20.2 & T.O. & 620.8 & 2880.0 & 180.0 & 565.1 & 210.8 & 1030.8 & 836.9 & 141.7 & 327.6 & 148.2 & $52.64\times$ & $1.52\times$\\
TIF012 & 28.3 & 19.6 & 16.0 & 6 & 3.0 & 36.0 & 16.1 & 4.9 & 69.2 & 6.5 & 2.9 & 2.7 & 3.6 & $0.55\times$ & $0.10\times$\\
TIF014 & 21.0 & 12.8 & T.O. & 3.8 & 8.0 & 2.0 & 32.3 & 122.7 & 52.6 & 24.3 & 8.1 & 5.6 & 51.8 & $7.00\times$ & $0.10\times$\\
XML001 & 10.7 & 3.4 & T.O. & 106.3 & 8.0 & 1200.0 & 1174.2 & 1034.3 & T.O. & 645.5 & 632.3 & 681.2 & 0.3 & $65.15\times$ & $0.03\times$\\
XML002 & 0.5 & 1.8 & 1.6 & T.O. & 10080.0 & T.O. & T.O. & 1419.9 & T.O. & T.O. & 1266.1 & 1225.9 & 788.3 & $3663.93\times$ & $3.20\times$\\
XML003 & 1.7 & 2.8 & 2.1 & T.O. & 59.0 & 300.0 & 596.6 & 792.4 & 1222.4 & 434.0 & 329.2 & 669.9 & 0.3 & $286.70\times$ & $0.18\times$\\
XML006 & 5.1 & T.O. & T.O. & T.O. & 10080.0 & T.O. & T.O. & T.O. & T.O. & T.O. & T.O. & T.O. & 1431.2 & $423.39\times$ & $280.63\times$\\
XML008 & 16.1 & T.O. & T.O. & T.O. & T.O. & T.O. & T.O. & T.O. & T.O. & T.O. & T.O. & T.O. & T.O. & $89.44\times$ & $89.44\times$\\
XML009 & 5.5 & 7.9 & T.O. & 41.3 & 13.0 & 60.0 & 5.9 & 27.4 & 69.5 & 6.0 & 60.4 & 118.2 & 52.8 & $28.82\times$ & $1.07\times$\\
XML010 & 17.4 & 8.5 & T.O. & T.O. & T.O. & T.O. & T.O. & T.O. & T.O. & T.O. & T.O. & T.O. & T.O. & $75.90\times$ & $0.49\times$\\
XML011 & 20.6 & T.O. & T.O. & T.O. & T.O. & T.O. & T.O. & T.O. & T.O. & T.O. & T.O. & T.O. & T.O. & $69.90\times$ & $69.90\times$\\
XML012 & 44.1 & T.O. & T.O. & 434.0 & 2880.0 & T.O. & 835.9 & 835.4 & T.O. & 960.5 & 1116.2 & 1294.7 & T.O. & $29.40\times$ & $9.84\times$\\
XML017 & 1.2 & 3.4 & 1.1 & 40.8 & 0.7 & 2.0 & 0.3 & 0.3 & 0.6 & 0.3 & 0.4 & 1.2 & 0.3 & $3.57\times$ & $0.25\times$\\
LUA001 & 1.1 & 2.1 & T.O. & T.O. & T.O. & T.O. & T.O. & T.O. & T.O. & T.O. & T.O. & T.O. & T.O. & $1200.16\times$ & $1.91\times$\\
LUA002 & 5.2 & 7.2 & 5.3 & T.O. & T.O. & T.O. & T.O. & T.O. & T.O. & T.O. & 1111.9 & T.O. & 1355.9 & $224.36\times$ & $1.02\times$\\
LUA003 & 0.8 & 5.5 & 1.5 & T.O. & T.O. & T.O. & T.O. & T.O. & T.O. & T.O. & T.O. & T.O. & 1309.7 & $1487.16\times$ & $1.88\times$\\
LUA004 & 15.3 & 4.0 & 2.7 & 6.3 & 23.0 & 180.0 & 205.4 & 253.3 & 983.3 & 105.6 & 970 & 876.9 & 672 & $23.33\times$ & $0.18\times$\\
PDF001 & 110.7 & T.O. & T.O. & T.O. & T.O. & T.O. & T.O. & T.O. & T.O. & T.O. & T.O. & T.O. & T.O. & $13.01\times$ & $13.01\times$\\
PDF002 & 0.8 & 2.9 & T.O. & T.O. & T.O. & T.O. & 1208.0 & 1353.1 & 1405.1 & 1348.7 & T.O. & T.O. & T.O. & $1603.94\times$ & $3.62\times$\\
PDF003 & 4.7 & 5.3 & 2.5 & 7.9 & 60.0 & 360.0 & 351.5 & 416.2 & 609.5 & 273.4 & 113.0 & 267.6 & 1156.4 & $64.24\times$ & $0.53\times$\\
PDF004 & 383.2 & T.O. & T.O. & T.O. & T.O. & T.O. & T.O. & T.O. & T.O. & T.O. & T.O. & T.O. & T.O. & $3.76\times$ & $3.76\times$\\
PDF005 & 5.3 & 12.4 & 7.9 & T.O. & T.O. & T.O. & T.O. & T.O. & T.O. & T.O. & T.O. & T.O. & T.O. & $226.73\times$ & $1.49\times$\\
PDF006 & 772.7 & T.O. & T.O. & 35.0 & 10080.0 & T.O. & T.O. & T.O. & T.O. & T.O. & 1432.0 & T.O. & 1334.9 & $2.63\times$ & $0.05\times$\\
PDF008 & 2.2 & 3.2 & 2.1 & T.O. & 10080.0 & T.O. & T.O. & T.O. & 1437.4 & 1375.9 & T.O. & 1419.8 & T.O. & $869.64\times$ & $0.95\times$\\
PDF009 & 0.9 & 3.1 & 3.1 & T.O. & T.O. & T.O. & T.O. & T.O. & T.O. & T.O. & T.O. & T.O. & T.O. & $1333.91\times$ & $3.44\times$\\
PDF010 & 6.5 & 8.4 & 4.2 & 384.7 & 28.0 & 120.0 & 26.9 & 9.1 & 7.8 & 15.2 & 222.6 & 272.4 & 57.5 & $14.83\times$ & $0.65\times$\\
PDF011 & 139.2 & T.O. & T.O. & T.O. & T.O. & 300.0 & T.O. & 1293.0 & 1232.4 & 1353.8 & 1116.9 & 970.4 & 1304.2 & $8.84\times$ & $2.16\times$\\
PDF012 & 1.9 & 5.6 & 3.9 & T.O. & T.O. & T.O. & T.O. & T.O. & T.O. & T.O. & T.O. & T.O. & T.O. & $632.00\times$ & $2.05\times$\\
PDF014 & 1313.3 & T.O. & T.O. & T.O. & T.O. & T.O. & T.O. & T.O. & T.O. & T.O. & T.O. & T.O. & T.O. & $1.10\times$ & $1.10\times$\\
PDF016 & 1.7 & 3.2 & 1.4 & 6.8 & 0.8 & 8.0 & 0.9 & 1.1 & 0.5 & 1.0 & 0.4 & 0.5 & 3.4 & $1.37\times$ & $0.24\times$\\
PDF018 & 0.9 & 3.1 & 1.7 & T.O. & 120.0 & 300.0 & 1094.7 & 11.3 & T.O. & 28.2 & 549.5 & 1328.4 & 31.7 & $587.83\times$ & $1.89\times$\\
PDF019 & 4.0 & 22.9 & T.O. & T.O. & T.O. & T.O. & 1396.2 & T.O. & 1182.4 & 1296.8 & 179.7 & 736.6 & T.O. & $280.30\times$ & $5.72\times$\\
PDF021 & 4.4 & 8.1 & 10.1 & T.O. & 10080.0 & 960.0 & T.O. & T.O. & 1402.3 & T.O. & 1205.4 & 912.7 & 1423.6 & $412.16\times$ & $1.84\times$\\
PDF022 & 1.3 & 3.1 & 2.0 & T.O. & T.O. & T.O. & T.O. & T.O. & T.O. & T.O. & T.O. & T.O. & T.O. & $923.40\times$ & $1.54\times$\\
SQL001 & 1.4 & T.O. & T.O. & T.O. & T.O. & T.O. & T.O. & T.O. & T.O. & 1371.0 & 1374.8 & T.O. & T.O. & $1020.58\times$ & $979.29\times$\\
SQL002 & 60.1 & T.O. & T.O. & 104.1 & 1.0 & 7.0 & 19.8 & 15.0 & T.O. & 16.9 & 3.3 & 9.8 & 46.4 & $6.30\times$ & $0.02\times$\\
SQL003 & 4.2 & 2.4 & T.O. & T.O. & 2880.0 & T.O. & T.O. & T.O. & T.O. & 1249.0 & 1282.1 & 1378.6 & T.O. & $334.76\times$ & $0.57\times$\\
SQL004 & 1.8 & T.O. & T.O. & T.O. & T.O. & T.O. & T.O. & 1337.1 & T.O. & T.O. & T.O. & T.O. & T.O. & $795.24\times$ & $742.83\times$\\
SQL005 & 20.7 & T.O. & T.O. & T.O. & T.O. & T.O. & 1165.1 & 927.8 & T.O. & 756.5 & 902 & 860.4 & T.O. & $59.15\times$ & $36.55\times$\\
SQL006 & 1.5 & T.O. & T.O. & T.O. & T.O. & T.O. & T.O. & T.O. & T.O. & T.O. & T.O. & T.O. & T.O. & $960.00\times$ & $960.00\times$\\
SQL007 & 6.2 & 5.7 & 2.7 & T.O. & T.O. & T.O. & T.O. & T.O. & T.O. & T.O. & T.O. & T.O. & T.O. & $193.66\times$ & $0.44\times$\\
SQL010 & 3.0 & 5.6 & T.O. & T.O. & T.O. & T.O. & T.O. & T.O. & T.O. & T.O. & T.O. & T.O. & T.O. & $440.16\times$ & $1.87\times$\\
SQL011 & 33.0 & T.O. & T.O. & T.O. & T.O. & T.O. & T.O. & T.O. & T.O. & T.O. & T.O. & T.O. & T.O. & $43.64\times$ & $43.64\times$\\
SQL012 & 4.5 & 2.0 & 2.7 & T.O. & 2880.0 & 1020.0 & 854.6 & 1105.6 & T.O. & 1232.6 & 1024 & 1104.6 & 1324.3 & $248.71\times$ & $0.44\times$\\
SQL013 & 5.8 & T.O. & T.O. & T.O. & 23.0 & 1260.0 & 1168 & T.O. & T.O. & 1004.1 & 1141.8 & 1293.4 & T.O. & $208.77\times$ & $3.97\times$\\
SQL014 & 14.5 & T.O. & T.O. & T.O. & 33.0 & 480.0 & 378.3 & 272.5 & T.O. & 325.8 & 30.5 & 23.7 & 254.1 & $43.44\times$ & $1.63\times$\\
SQL015 & 3.0 & 8.2 & 3.7 & T.O. & 2880.0 & T.O. & 1387.2 & 1375.1 & T.O. & T.O. & 1092.7 & 1224.9 & T.O. & $421.44\times$ & $1.23\times$\\
SQL018 & 7.6 & 3.5 & T.O. & 9.8 & 1.0 & 8.0 & 27.3 & 33.4 & T.O. & 30.1 & 20.5 & 12.2 & 212.7 & $35.51\times$ & $0.13\times$\\
SQL020 & 0.6 & 2.1 & 2.4 & T.O. & 60.0 & 720.0 & 615.2 & 1322.6 & T.O. & 944.3 & 732.1 & 903.7 & 1298.2 & $1316.75\times$ & $3.50\times$\\
SSL001 & 10.8 & T.O. & T.O. & 180.8 & T.O. & 360.0 & 852.5 & N.A. & 24.0 & 288.8 & 487.1 & 524.5 & 802.0 & $65.99\times$ & $2.22\times$\\
SSL002 & 5.5 & 4.0 & 7.3 & 11.3 & 33.0 & T.O. & 2.5 & N.A. & T.O. & 1.4 & 0.6 & 1.4 & 0.5 & $48.63\times$ & $0.09\times$\\
SSL003 & 0.9 & 7.7 & 1.9 & 5.8 & 0.2 & 0.5 & 2.9 & N.A. & T.O. & 1.8 & 0.7 & 1.8 & 0.3 & $147.84\times$ & $0.22\times$\\
SSL006 & T.O. & 24.0 & T.O. & T.O. & T.O. & T.O. & T.O. & N.A. & T.O. & T.O. & T.O. & T.O. & T.O. & -- & --\\
SSL009 & 3.2 & T.O. & T.O. & T.O. & T.O. & T.O. & 1098.6 & N.A. & T.O. & 477.5 & 158.7 & 1208.3 & 0.7 & $329.09\times$ & $0.22\times$\\
SSL016 & 33.7 & T.O. & T.O. & T.O. & T.O. & T.O. & T.O. & N.A. & T.O. & T.O. & T.O. & T.O. & T.O. & $42.73\times$ & $42.73\times$\\
SSL020 & T.O. & 9.1 & T.O. & T.O. & 10080.0 & T.O. & 1327.4 & N.A. & T.O. & 842.2 & 254.2 & 538.1 & 555.3 & -- & --\\
PHP001 & 1.8 & 11.0 & 9.8 & T.O. & T.O. & T.O. & T.O. & N.A. & N.A. & T.O. & T.O. & T.O. & T.O. & $641.16\times$ & $5.44\times$\\
PHP002 & 1.1 & T.O. & T.O. & 12.3 & T.O. & T.O. & 0.3 & N.A. & N.A. & 0.3 & 0.3 & 0.3 & 0.3 & $524.89\times$ & $0.27\times$\\
PHP003 & T.O. & 12.3 & T.O. & T.O. & T.O. & T.O. & T.O. & N.A. & N.A. & T.O. & T.O. & T.O. & T.O. & -- & --\\
PHP004 & 2.4 & 10.0 & 4.0 & T.O. & 10.0 & 360.0 & 7.8 & N.A. & N.A. & 45.4 & 4.3 & 12.8 & 629.3 & $105.15\times$ & $1.67\times$\\
PHP005 & 2.1 & T.O. & T.O. & T.O. & T.O. & T.O. & T.O. & N.A. & N.A. & T.O. & T.O. & T.O. & T.O. & $685.71\times$ & $685.71\times$\\
PHP009 & 2.7 & 15.4 & 12.9 & 13.6 & 2.0 & 840.0 & 477.2 & N.A. & N.A. & 1058.5 & 384.2 & 357.1 & 0.3 & $117.08\times$ & $0.11\times$\\
PHP010 & T.O. & 19.6 & T.O. & T.O. & T.O. & T.O. & T.O. & N.A. & N.A. & T.O. & T.O. & T.O. & T.O. & -- & --\\
PHP011 & 3.3 & 10.2 & T.O. & T.O. & 1.0 & 5.0 & T.O. & N.A. & N.A. & T.O. & T.O. & T.O. & T.O. & $305.95\times$ & $0.30\times$\\
PHP013 & 1.8 & 6.7 & 3.0 & T.O. & T.O. & T.O. & T.O. & N.A. & N.A. & T.O. & T.O. & T.O. & T.O. & $640.54\times$ & $1.67\times$\\
PHP014 & 0.8 & 3.9 & 3.5 & T.O. & T.O. & T.O. & T.O. & N.A. & N.A. & T.O. & T.O. & T.O. & T.O. & $1440.92\times$ & $4.38\times$\\
    \midrule
        \textbf{CVE Cov.} & \textbf{80} & \textbf{57}(-23) & \textbf{38}(-42) & \textbf{28}(-52) & \textbf{46}(-34) & \textbf{41}(-39) & \textbf{45}(-35) & \textbf{37}(-43) & \textbf{29}(-51) & \textbf{48}(-32) & \textbf{52}(-28) & \textbf{51}(-29) & \textbf{48}(-32) & -- & --\\
\bottomrule
\end{tabularx}
\par\vspace{2pt}
\renewcommand{\arraystretch}{1}
}
\end{table*}

\subsection{Oracle Synthesis Case Studies}
\label{sec:oracle_case_studies}
To complement the quantitative evaluation, we present representative case studies
illustrating DIG's oracle synthesis under different categories.
We include examples corresponding to \textbf{L1 (Syntactic Equivalence)},
\textbf{L2 (Semantic Equivalence)}, \textbf{L3 (Sufficient-Condition Correctness)},
\textbf{Synthesis Errors}, and \textbf{Reference Errors}.

\subsubsection{L1 (Syntactic Equivalence)}
In this case, a synthesized oracle is considered correct if
it is syntactically identical to the ground-truth oracle. 
CVE-2019-14494 (PDF001), shown in Figure~\ref{fig:pdf001_vuln_code}, 
is an example where the synthesized oracle is identical to the ground-truth oracle,
which is \texttt{O(surface\_width == 0 || surface\_height == 0)}.

\subsubsection{L2 (Semantic Equivalence)}
In this case, a synthesized oracle is considered correct if it is semantically
equivalent to the ground-truth oracle, even though the two predicates are syntactically different.
CVE-2016-1836 (XML012) is an example where the synthesized oracle is 
\texttt{O(ctxt->input->base != initial\_base)} while the ground-truth oracle is 
\texttt{O((end - len) != (BASE\_PTR + startPosition))}.
Although the two predicates differ syntactically, they are semantically equivalent
because both compare the current buffer pointer with its initial base pointer.
The expressions \texttt{(end - len)} and \texttt{(BASE\_PTR + startPosition)} correspond to these two pointers,
so both oracles characterize the same condition.

\subsubsection{L3 (Sufficient-Condition Correctness)}
In this case, a synthesized oracle is considered correct if it captures a
sufficient condition for the ground-truth oracle. CVE-2019-9200 (PDF007) is
an example where the synthesized oracle is \texttt{O(readChars == -1)}, while
the ground-truth oracle is \texttt{O(readChars < 0)}. 
Here, \texttt{readChars == -1} implies \texttt{readChars < 0}, so the synthesized
oracle captures a valid but more restrictive condition.

\subsubsection{Synthesis Errors}
In this case, a synthesized oracle is considered incorrect with respect to the ground-truth oracle. 
For example, in CVE-2017-11362 (PHP013), the ground-truth oracle checks \texttt{slocale\_len > INTL\_MAX\_LOCALE\_LEN}, where 
\texttt{INTL\_MAX\_LOCALE\_LEN} is 255. However, the synthesized oracle incorrectly infers this macro value
and checks \texttt{slocale\_len > 80} instead. 
As a result, the synthesized oracle characterizes a different condition 
from the ground-truth oracle and is classified as a synthesis error. 
This differs from Sufficient-Condition Correctness, where the synthesized oracle must be a 
sufficient condition for the ground-truth oracle. 
Here, \texttt{slocale\_len > 80} is not a sufficient condition for \texttt{slocale\_len > 255}.

\subsubsection{Reference Errors}
In this case, the Magma reference oracle itself is incorrect with respect to the ground-truth vulnerability condition. 
For example, in CVE-2016-10270 (TIF007), the reference oracle uses the predicate 
\texttt{nstrips > TIFFhowmany\_32(td->td\_imagelength, rowsperstrip)}, whereas the correct condition
is \texttt{nstrips32 < TIFFhowmany\_32(td->td\_imagelength, rowsperstrip)}. 
This discrepancy indicates that the reference oracle characterizes an incorrect condition,
and we therefore classify this case as a reference error.

\begin{figure}[H]
    \centering
    \adjustbox{width=\linewidth}{
    \begin{minipage}{\linewidth}
    \begin{Verbatim}[fontsize=\footnotesize, breaklines=true, breakanywhere=false]
// poppler/SplashOutputDev.cc:4220-4299 

// Semantic constraint: XStep/YStep must match BBox dimensions
const double *bbox = tPat->getBBox();
width = bbox[2] - bbox[0];   // BBox width
height = bbox[3] - bbox[1];  // BBox height
if (xStep != width || yStep != height)
    return false;  // PoC: both are 1e-340
// ... (lines 4222-4245: CTM processing and initial transform)
// Step 1: Floating-point underflow
result_width = (int)ceil(fabs(kx * width * (x1 - x0)));
// kx=1e-5 (from CTM), width=1e-340 (from BBox), (x1-x0)=1 (clip)
// -> 1e-5 x 1e-340 x 1 = 1e-345 < 4.94e-324 (IEEE 754 min subnormal)
// -> underflows to 0.0 -> result_width = 0
result_height = (int)ceil(fabs(ky * height * (y1 - y0)));
// ... (lines 4248-4259: matrix setup with DPI and pattern matrix)
// Step 2: Zero propagation through scaling factor
sx = (double)result_width / (surface_width * (x1 - x0));
// sx = 0 / (non-zero) = 0
sy = (double)result_height / (surface_height * (y1 - y0));
m1.m[0] *= sx;  // m1.m[0] becomes 0
m1.m[3] *= sy;
m1.transform(width, height, &kx, &ky);
// kx = 0 x width + 0 x height + 0 = 0
// Step 3: Branch selection (small tile path)
if (fabs(kx) < 1 && fabs(ky) < 1) {
    // ... (lines 4267-4275: special handling for tiny tiles)
} else {
    // ... (lines 4277-4286: size limit checks)
    // Step 4: Final surface_width computation
    surface_width = (int)ceil(fabs(kx)); // kx = 0 -> surface_width = 0
    surface_height = (int)ceil(fabs(ky));   
    // Step 5: DIG's oracle: Division by zero
    O(surface_width == 0 || surface_height == 0); 
    repeatX = result_width / surface_width;
    repeatY = result_height / surface_height;
}
    \end{Verbatim}
    \end{minipage}
    }
    \caption{CVE-2019-14494: a division-by-zero error caused by floating-point underflow.}
    \label{fig:pdf001_vuln_code}
\end{figure}

\begin{figure}[H]
    \centering
    \adjustbox{width=\linewidth}{
    \begin{minipage}{\linewidth}
    \begin{Verbatim}[fontsize=\footnotesize, breaklines=true, breakanywhere=false]
// poppler/Gfx.cc:2001-2005 (Gfx::doTilingPatternFill)
det = ctm[0] * ctm[3] - ctm[1] * ctm[2];
if (fabs(det) < 0.000001) {  // 1e-6 threshold
    error(errSyntaxError, getPos(), "Singular matrix in tiling pattern fill");
    return;
}
    \end{Verbatim}
    \end{minipage}
    }
    \caption{Singularity check in \texttt{Gfx::doTilingPatternFill()} for 
    CVE-2019-14494.}
    \label{fig:pdf001_singularity_check}
\end{figure}

\begin{figure}[H]
    \centering
    \adjustbox{width=\linewidth}{
    \begin{minipage}{\linewidth}
    \begin{Verbatim}[fontsize=\footnotesize, breaklines=true, breakanywhere=false]
// poppler/SplashOutputDev.cc (patched version)
surface_width = (int) ceil (fabs(kx));
surface_height = (int) ceil (fabs(ky));
// adjust repeat values to completely fill region
+ if (unlikely(surface_width == 0 || surface_height == 0)) {
+    state->setCTM(savedCTM[0], savedCTM[1], savedCTM[2], savedCTM[3], savedCTM[4], savedCTM[5]);
+    return false;
+ }
repeatX = result_width / surface_width;
repeatY = result_height / surface_height;
if (surface_width * repeatX < result_width)
    \end{Verbatim}
    \end{minipage}
    }
    \caption{Patch for CVE-2019-14494 that adds a zero-check before division.}
    \label{fig:pdf001_patch}
\end{figure}

\begin{figure}[H]
    \centering
    \adjustbox{width=\linewidth}{
    \begin{minipage}{\linewidth}
    \begin{Verbatim}[fontsize=\footnotesize, breaklines=true, breakanywhere=false]
1 0 obj << /Type /Catalog /Pages 2 0 R >> endobj
2 0 obj << /Type /Pages /Kids [3 0 R] /Count 1 >> endobj
3 0 obj << /Type /Page /Parent 2 0 R /MediaBox [0 0 200 200]
   /Resources << /Pattern << /P1 5 0 R >> >> /Contents 4 0 R >> endobj
4 0 obj << /Length 77 >> stream
q
0 0 1 1 re W n              %
0.00001 0 0 1 0 0 cm        %
/Pattern cs /P1 scn
0 0 200 200 re f            %
Q
endstream endobj
5 0 obj << /Type /Pattern /PatternType 1 /PaintType 1 /TilingType 2
   /BBox [0 0 1e-340 16]    %
   /XStep 1e-340 /YStep 16  %
   /Matrix [1 0 0 1 0 0] /Resources << >> /Length 25 >> stream
1 0 0 rg 0 0 1e-340 16 re f
endstream endobj
    \end{Verbatim}
    \end{minipage}
    }
    \caption{Complete PoC for CVE-2019-14494.}
    \label{fig:pdf001_poc}
\end{figure}

\begingroup
\makeatletter
\setlength{\@fptop}{0pt}
\setlength{\@fpbot}{0pt}
\makeatother
\begin{figure}[!t]
    \centering
    \adjustbox{width=\linewidth}{
    \begin{minipage}{\linewidth}
    \begin{Verbatim}[fontsize=\footnotesize, breaklines=true, breakanywhere=false]
// tif_next.c  NeXTDecode()
#define SETPIXEL(op, v) {
    switch (npixels++ & 3) {
    case 0: op[0]  = (unsigned char)((v) << 6); break;
    case 1: op[0] |= (v) << 4; break;
    case 2: op[0] |= (v) << 2; break;
    case 3: *op++ |= (v); op_offset++; break; // advances op 1 byte every 4 pixels
    }}

static int NeXTDecode(TIFF* tif, uint8_t* buf, tmsize_t occ, uint16_t s) {
    tmsize_t scanline = tif->tif_scanlinesize; // from td_imagewidth via TIFFScanlineSize
    ...
    for (row = buf; cc > 0 && occ > 0; occ -= scanline, row += scanline) {
        // row points to current output row; allocated size = scanline bytes
        n = *bp++; cc--;
        switch (n) {
        // ... LITERALROW / LITERALSPAN cases omitted ...
        default: {
            uint32_t npixels = 0, grey;
            tmsize_t op_offset = 0;
            uint32_t imagewidth = tif->tif_dir.td_imagewidth;
            if (isTiled(tif))
                imagewidth = tif->tif_dir.td_tilewidth; // C5: tilewidth overrides
            op = row;
            for (;;) {
                grey = (uint32_t)((n >> 6) & 0x3);  n &= 0x3f;
                while (n-- > 0 && npixels < imagewidth) {
                    O(op_offset >= scanline);   // oracle: op past row boundary
                    SETPIXEL(op, grey);         // OOB write past row[scanline-1]
                }
                if (npixels >= imagewidth) break;
                n = *bp++; cc--;
            }
        }}
    }
}

// tif_strip.c  TIFFScanlineSize64() — sets tif->tif_scanlinesize
uint64_t TIFFScanlineSize64(TIFF* tif) {
    TIFFDirectory *td = &tif->tif_dir;
    // ... YCbCr path omitted; NeXT takes the contig non-YCbCr branch:
    uint64_t scanline_samples;
    scanline_samples = _TIFFMultiply64(tif,
        td->td_imagewidth, td->td_samplesperpixel, module);
    // scanline_size depends solely on td_imagewidth
    scanline_size = TIFFhowmany_64(
        _TIFFMultiply64(tif, scanline_samples,
                        td->td_bitspersample, module), 8);
    return (scanline_size);
}
    \end{Verbatim}
    \end{minipage}
    }
    \caption{CVE-2015-8784 (TIF008): heap buffer over-write in
    \texttt{NeXTDecode} (\texttt{tif\_next.c}).
    \texttt{SETPIXEL} writes past the scanline buffer when
    \texttt{op\_offset} $\geq$ \texttt{scanline}, triggered via a tiled TIFF
    where \texttt{td\_tilewidth} $>$ \texttt{td\_imagewidth}.}
    \label{fig:tif008_vuln_code}
\end{figure}
\endgroup

\begin{figure}[H]
    \centering
    \begin{tcolorbox}[colback=gray!5, colframe=gray!40, width=\linewidth, title=Prompt template for Cursor agent, fonttitle=\bfseries]
    \small
    \textbf{\#\# Mission}\\
    You are a security expert who finds PoC input for bugs in C/C++ programs.\\
    Your goal is to find a PoC input that triggers the bug predicate.

    \medskip
    \textbf{\#\# System Input}\\
    You will receive structured input containing:
    \begin{enumerate}[leftmargin=*, itemsep=1pt, topsep=2pt]
        \item \textbf{ENTRY\_SOURCE\_CODE}: Entry functions (main/FuzzerTestOneInput) and target location functions with line numbers
        \item \textbf{TARGET\_LOCATIONS}: Bug predicate locations with format `loc=file:line, code=source\_line'
    \end{enumerate}
    Use this information to analyze the program, locate the bug predicate, and write a Python script that generates a PoC input which triggers the bug at the target location.
    \end{tcolorbox}
    \caption{Prompt template for Cursor agent.}
    \label{fig:cursor_initial_system_prompt}
\end{figure}

\end{document}